\documentclass{article}
\usepackage{amsfonts,amssymb,latexsym,amsmath,enumerate,amsthm,cite}
\newcommand{\mytitle}[1]{\large \sc #1 \\}
\newcommand{\avtor}[1]  {\large \it #1 \\}
\newcommand{\HAEH } {Ehrenfest }

\makeatother 

\def\pa{\partial}
\def\a{\alpha}
\def\va{\varphi}
\def\vk{\varkappa}
\def\tvk{\tilde\varkappa}

\def\iy{\infty}
\def\De{\Delta}
\def\de{\delta}

\def\FF{{\mathfrak F}}
\def\FH{{\mathfrak H}}
\def\fg{{\mathfrak g}}

\def\CD{{\mathcal D}}
\def\CH{{\mathcal H}}

\def\CP{{\mathcal P}}

\def\BR{{\mathbb R}}

\def\BS{{\mathbb S}}
\def\BZ{{\mathbb Z}}
\def\CF{{\mathcal F}}

\def\lan{\langle}
\def\ran{\rangle}
\def\Im{\mathop{\rm Im}\nolimits}

\def\Im{\mathop{\rm Im}\nolimits}
\def\Re{\mathop{\rm Re}\nolimits}
\def\Sp{\mathop{\rm Sp}\nolimits\,}

\def\const{\mathop{\rm const}\nolimits}

\def\h{\hbar}
\def\le{\leqslant}
\def\ge{\geqslant}
\def\dac{\displaystyle\frac}
\def\inl{\int\limits}

\def\dil{\displaystyle\int\limits}

\def\dum{\displaystyle\sum}

\def\le{\leqslant}
\def\ge{\geqslant}

\def\{{\lbrace}
\def\}{\rbrace}

\def\thesection{\arabic{section}.}
\def\theequation{\thesection\arabic{equation}}

\makeatletter \@addtoreset{equation}{section} \makeatother

\newtheorem{assum}{Assumption}%

\newtheorem{lem}{Lemma}

\newtheorem{demo}{Statement}

\def\wh{\widehat}

\makeatletter \@addtoreset{teo}{section}
\@addtoreset{lem}{section} \@addtoreset{samp}{section}
\@addtoreset{chap}{part} \@addtoreset{section}{part}
\@addtoreset{dmd}{section} \@addtoreset{demo}{section}
\@addtoreset{sled}{teo} \makeatother \makeindex

\begin{document}

\begin{center}

\mytitle{The \HAEH system and the rest point spectrum for a Hartree-type Equation}

\bigskip
\avtor{V.~V.~Belov,$^{\dag}$\footnote{e-mail: belov@amath.msk.ru}
M.~F.~Kondratieva,$^{\#}$\footnote{e-mail: mkondra@math.mun.ca}
A.~Yu.~Trifonov$^*$\footnote{e-mail: trifonov@mph.phtd.tpu.edu.ru}}
\bigskip

$^{\dag}$ {\it Department of Applied Mathematics,\\
Moscow State Institute of Electronics and Mathematics,\\
Trehsvjatitelsky per., 3/12, Moscow  109028, Russia}\vskip 0.2cm
$^{\#}$ {\it Department of Mathematics and Statistics,\\
Memorial University of Newfoundland,\\
St. John's A1C 5S7, Canada} \vskip 0.2cm
$^{*}$ {\it Laboratory of Mathematical Physics, \\
Mathematical Physics Department,\\
Tomsk Polytechnical University,\\
Lenin ave. 30, Tomsk 634050, Russia}
\end{center}

\noindent {\small {\bf Abstract.} Following Ehrenfest's approach,
the problem of quantum-classical correspondence  can be treated in
the class of trajectory-coherent functions that approximate as
$\h\to0$ a quantum-mechanical state. This idea leads to a family
of systems of ordinary differential equations, called \HAEH
$M$-systems ($M=0,1,2,\dots$), formally equivalent to the
semiclassical approximation for the linear Schr\"odinger equation
\cite{69,Bagre}.

In this paper a similar approach is undertaken for a nonlinear Hartree-type equation
with a smooth integral kernel.
It is demonstrated how quantum characteristics can be retrieved directly from the corresponding \HAEH systems,
without solving the quantum equation: the semiclassical asymptotics for the spectrum are obtained from the rest point solution.
One of the key steps is derivation  of a modified nonlinear superposition principle valid in the class of
trajectory-coherent
quantum states.
}

\section*{Introduction}

Semiclassical methods play a distinguished role among asymptotic
approaches in linear mathematical physics. From the very beginning
of quantum mechanics semiclassical approximation has been one of
the the main technical tools to address its two aspects: pragmatic
and philosophical.

The pragmatic (computational)  aspect relies on the presence of a
{\it small} parameter $\hbar$ as a factor next to the derivatives.
The pattern is demonstrated in the  Schr\"odinger evolution
equation
\begin{equation}
i\h\dac{\pa\Psi}{\pa t}=\hat \CH\Psi,\qquad \hat\CH=
\frac{\hat{\vec
p}{\,}^2}{2m}+U(\vec x),
\quad \hat{\vec p} = -i\h \nabla_x,\quad \vec x\in\BR^n,\label{0.1}
\end{equation}
corresponding to the classical 
Hamilton function
\begin{equation}
\CH(\vec p, \vec
x,t) =\frac{\vec
p{\,}^2}{2m}+U(\vec x). \label{0.2}
\end{equation}
While Planck's $\hbar$ is a dimensional constant,
there exists a large class of quantum-mechanical problems where a small dimensionless
parameter, proportional to $\hbar$, is present.  Accordingly,
there is a mathematical problem to construct an approximate
(with respect to that
parameter) solution of the quantum mechanical equation. Such an approximate
solution  is traditionally termed
{\it the semiclassical asymptotics as $\hbar\to 0$}.

The philosophical aspect is related to the correspondence
principle, one of the cornerstones of quantum mechanics. Despite
the fact that quantum mechanics in its axiomatic formalization is
a self-consistent theory and does not appeal to the classical
mechanics, the correspondence principle requires the classical
equations of motion to emerge from the quantum theory in the limit
$\hbar\to 0$.

Obviously, there is no universal (i.e. physical problem
independent) way to obtain arbitrary classical values from
quantum-mechanical values. In each particular case it is necessary
to specify in what sense a  quantum characteristic becomes
classical as $\hbar \to 0$.  The problem of deriving classical
equations of motion from those of quantum mechanics in the limit
$\hbar \to 0$ is one of the principal questions of the
quantum-classical correspondence.

Historically, there are a number of approaches to the problem. One
of them is due to Born \cite{born}, in which a quantum system is
approximately described by the classical statistical ensemble
expressed via a semiclassical wave function. A justification of
this approach is based on the  construction of a semiclassical
solution to the quantum equation.  The time-global version of such
a construction is known as   the  Maslov canonical operator
\cite{1, MasFed}. In this approach the correspondence principle
reveals itself in the fact that the principal term of the
asymptotic expansion of the quantum density matrix is a solution
of the classical Liouville equation.

Another approach, suggested by Ehrenfest \cite{Ehrenfest},  is
based on the idea that Newtonian equations of motion can be
obtained in the limit $\hbar \to 0$ from equations for mean values
of the corresponding quantum-mechanical observables. More
generally, any ordinary differential equations (ODE) obtained in
the same manner from equations of quantum mechanics can be called
classical. The correspondence between a quantum observable and its
classical counterpart (assuming that such exists) is understood as
follows: the quantum mean value $\lan \hat A\ran_\Psi$ of the
observable $\hat A=A(\hat z,\h)$ calculated with respect to some
special non-stationary states $\Psi(t;\h)$ must yield in the limit
$\hbar\to 0$ the corresponding classical observable $A$ evaluated on a
certain classical trajectory $z(t)$ in the phase space
\begin{equation}
\label{Eh} \lim_{\h\to 0} \lan \hat A\ran_\Psi = A(z(t),0).
\end{equation}

For the Schr\"odinger equation  (\ref{0.1}) the Ehrenfest approach relies on 
  states $\Psi(\vec x,t;\h) $ that are
{\it localized  on the classical trajectory} in the following sense: the mean values
\begin{equation}
\label{xpbar}
\bar{x}_k(t,\h)\equiv\langle\hat{ x}_k\rangle_\Psi=
\int_{\BR^n}x_k|\Psi|^2d\vec x,\quad
\bar{ p}_k(t,\h)\equiv \langle\hat{p}_k\rangle_\Psi=\int_{\BR^n}\Psi^*\hat p_k
\Psi d\vec x,\qquad k=\overline {1, n},
\end{equation}
of the operators of coordinates
$\hat{\vec x}=(x_1,\dots,x_n)$ and
momenta $\hat{\vec p}=-i\hbar\nabla$
calculated with respect to such states $\Psi(x,t;\h)$ in the limit $\hbar\to 0$
\begin{equation}
X_k(t)= \lim_{h\to 0}\bar{ x}_k(t,\h), \quad P_k(t)= \lim_{h\to
0}\bar{ p}_k(t,\h),\qquad k=\overline {1, n}, \label{0.14}
\end{equation}
obey the classical Hamiltonian system
\begin{equation}
m\dot {\vec X} =\vec P,\quad \dot{\vec  P} = -\nabla_x U(\vec X).
\label{0.4}
\end{equation}

A function $\Psi$ for which the limits (\ref{0.14}) exist was called in \cite{27,28} a {\it trajectory-coherent state}.

The technical implementation of the Ehrenfest  approach is based
on the construction of either exact or approximate
trajectory-coherent solutions of the Schr\"odinger equation. Exact
trajectory-coherent solutions are available only for special
Hamiltonians,   such as (\ref{0.1})  with quadratic potential.
Examples are well-known coherent and squeezed coherent states
\cite{Manko,Perel}. An approximate ($\h\to 0$) trajectory-coherent
solution, called a {\it semiclassically concentrated state} can be
constructed in a much wider class of problems, employing the ideas
of the complex WKB-Maslov method \cite{Maslov2,BeD2} (see also
\cite{45,46,57,42}). The correspondence principle is manifested in
this construction:
 a trajectory-coherent state is an approximate ($\h\to 0$) solution of (\ref{0.1})
 (i.e. it is a semiclassically concentrated state)  if and only if the trajectory (\ref{0.14})
satisfies the classical equation (\ref{0.4}). The semiclassically
concentrated states were first found for particles moving in a
potential field  \cite{47}, and later in an arbitrary
electromagnetic field  \cite{27,28}. Detailed bibliography can be
found in the reviews  \cite{42, Bagre}.

It was found that semiclassically concentrated states exist for
linear equations of quantum mechanics describing a charged
particle with spin or isospin  in an external field. In \cite{33,
37, 4.15,38,33a} the semiclassically concentrated states were
constructed for the Klein-Gordon and Dirac-Pauli equations in an
arbitrary electromagnetic field as well as for the Schr\"odinger
and Dirac equations in an arbitrary non-abelian field with gauge
group $SU(2)$.

The existence of the semiclassically concentrated states is essential for
the approach employed in this paper,  which  consists of the following.
Consider an observable $\hat A=A(\hat z)$ whose classical analog is $A(z)$. Its mean value in a
semiclassically concentrated state can be expressed to any accuracy  $O(\hbar^{(M+1)/2})$ via a solution
$\{ z(t),\Delta^2(t),\dots, \Delta^M(t)\}$
of a finite system of ODEs ,
\begin{equation}
\label{Napr}
\lan \hat A\ran_\Psi=A( z(t))+\sum\limits_{k=2}^M\,A_k( z(t))\cdot\Delta^k(t)+O(\hbar^{(M+1)/2}),
\qquad \Delta^k(t)=O(\hbar^{k/2}),
\end{equation}
where tensors  $A_k(z)$  comprise all partial derivatives of $A$ of order $k$ at the point $z$,
and tensors $\Delta^k$ comprise all moments of order $k$
(see (\ref{Del}) below).

The dimension of the ODE system  is determined by the order of accuracy $M$.
For instance, if $M=0$ or $1$ (it appears that $\Delta^1=0$ by construction), we obtain
 $$\lan \hat A\ran_\Psi=A( z(t))+O(\h),$$ where $ z(t)$ is subject to classical-mechanics equations,
in accordance to Ehrenfest's original idea (\ref{Eh}). In our
approach, by classical equations (of order $M\ge 0$) corresponding
to a quantum equation we mean that finite system of
ODEs whose solution provides accuracy $O(\hbar^{(M+1)/2})$ in (\ref{Napr}), and we call it 
 the {\it Ehrenfest $M$-system}.

The  Ehrenfest systems of finite order  are  truncations of an infinite ODE system, which
describes evolution of the mean values
for a basic infinite set of observables. For
the Schr\"odinger equation (\ref{0.1}) the basic set consists of $\hat {\vec x}$, $\hat {\vec p}$ and
a special basis of the universal enveloping
of the Heisenberg-Weyl algebra with
generators  $\hat{I}$, $\Delta\hat{x}_k=\hat{x}_k-\bar{x}_k(t)$,
$\Delta\hat{p}_k=\hat{p_k}-\bar{p}_k$, $1\le k\le n$,  where $\hat{I}$ is the
identity operator and  $\bar x_k$, $\bar p_k$ are defined in (\ref{xpbar}). The truncations
leading to $M$-systems are made due to the estimates for $\Delta^k$ in (\ref{Napr}),
which allow to disregard within given accuracy $O(\h^{(M+1)/2})$ all variables $\Delta^k$ for $k>M$
when the means are calculated with respect to semiclassically-concentrated states.
The $M=0$  truncation is simply Newton's system (\ref{0.4}); similarly, the
Ehrenfest 0-system for the Klein-Gordon equation is the Lorentz equation.

The  infinite ODE system for (\ref{0.1}) was derived in
\cite{69,88a,88} and it was called the Hamilton-Ehrenfest system
in \cite{Bagre}. The name reflects a non-trivial fact that the
infinite  system can be written in the  Hamiltonian form with
respect to a degenerate nonlinear Dirac bracket \cite{bk1}. A
system with similar algebraic structure was also derived for the
(matrix) Pauli equation \cite{bk2}. A Hamiltonian structure with a
degenerate Poisson bracket is also known for the $M=2$ truncation
\cite{bk1,Sadov96,KondraSad04}. The truncated systems were
independently introduced in \cite{Bal98} and used to study quantum
problems with underlying classically chaotic dynamics.

In a number of examples this approach was shown to agree with
known ``classical'' equations of motion even in the cases where no
corresponding classical observables existed. For the Dirac-Pauli
equation in an external field the \HAEH 0-system is a pair of
classical equations which are the Lorentz equation and the
Bargmann-Michel-Telegdi \cite{BMT} equation in which the field is
calculated on the trajectories of the Lorentz equation. The order
$M=2$ truncation obtained in \cite{33a} is a Frenkel type
\cite{frenkel1} ODE for spin motion. For  the Schr\"odinger and
Dirac equations in external fields with gauge group $SU(2)$ the
\HAEH $2$-system \cite{37,38} yields the Wong \cite{Wong} equation
for a non-abelian particle with isospin 1/2. More examples of
derivations of known ``classical'' equations from the
Dirac equation with external fields 
and the Prock equation are given in \cite{35,34,36}.

The \HAEH $M$-system is semiclassically  equivalent  with accuracy
$O(\hbar^{(M+1)/2})$ to the \newline Schr\"odinger equation in the
class of trajectory-concentrated states in the sense that it
allows us to calculate the  mean value of an observable directly
from the solutions of the system. An explicit formula for the
state is not required. It was also observed that under certain
conditions one  can obtain  asymptotics for pure quantum
characteristics, such as energy spectrum series, from stationary
or periodic solutions of the \HAEH 2-system.

The goal of this paper is to generalize the  approach for the case of a
nonlinear Hartree-type equation. In particular, we consider the following
equation of self-consistent field
\begin{equation}
\begin{aligned}
&i\h\frac{\pa\Psi}{\pa t}=\widehat{\CH}_\varkappa(\Psi)\Psi,  \\
&\widehat{\CH}_\varkappa(\Psi)=-\frac{\h^2}{2m}\nabla_x^2 +U(\vec x)+  \varkappa\int\limits_{\BR^n}V(\vec
x,\vec y)\bigl|\Psi(\vec y,t)\bigr|^2\,d \vec
y,\qquad \vec
x\in\BR^n, 
\end{aligned}\label{0.5}
\end{equation}
where $U(\vec x)$ and $V(\vec x,\vec y)$ are given smooth
potentials of the external electromagnetic field and the
self-consistent field respectively, and $\varkappa$ is a constant.

There are at least two reasons why the problem of quantum-classical
correspondence was not considered
in the spirit of Ehrenfest's approach, neither for a
nonlinear self-consistent field (\ref{0.5}), nor for more general
Hartree-type equations.

First, the operator $\widehat{\CH}_\varkappa(\Psi)$ does not have a natural classical
analog in the traditional sense, thus it is not {\it a priori}\ obvious which
dynamical $\Psi$-independent system is an appropriate candidate for the ``classical''
system in the limit $\hbar \to 0$.

Second, it is not clear whether the nonlinear quantum equation has either
exact or
approximate (as $\hbar \to 0$)  solutions that are trajectory-coherent
in the sense of (\ref{0.14}).

In the framework of our approach, a solution of the correspondence
problem includes three stages.

\begin{enumerate}
\item First, for an arbitrary one parameter family of phase space
trajectories  $Z(t,\h)=(\vec P(t,\h),\vec X(t,\h))$,
$t\in\mathbb{R}$, we introduce a class $\mathcal{P}_\h^t(Z(t,\h))$
of {\it trajectory-coherent functions}.  Exact construction is
given in Sect.~2.

Let $\alpha,\beta\in\BZ_+^n$ be multi-indices,  $|\alpha|=\sum_{k=1}^n \alpha_k$,
and $\vec X^\alpha=\prod_{j=1}^n X_j^{\alpha_j}$.
Let $\widehat{\Delta}_{\alpha\beta}$ be an operator  with Weyl
symbol  ${\Delta}_{\alpha\beta}(\vec p,\vec x)= \bigl({\vec
x}-{\vec X}(t)\bigr)^\alpha\bigl({\vec p}-{\vec
P}(t)\bigr)^\beta$.
The following property is essential for further steps:
the centered moments
\begin{equation}
\label{Del} \Delta_{\alpha\beta}(t,\h)=
\Bigl<\widehat{\Delta}_{\alpha\beta}\Bigr>_\Psi, \quad
\end{equation}
 calculated with respect to functions from class
$\mathcal{P}_\h^t(Z(t,\h))$ satisfy the  estimate
$$\Delta_{\alpha\beta}=O\Bigl(\h^{(|\alpha|+|\beta|)/2}\Bigr), \quad \hbar\to
0.
$$
Consequently, $k$-th order moments (i.e.\ those with $|\alpha|+|\beta|=k$)
are $O(\h^{k/2})$.

\item  Next, we assume that
equation (\ref{0.5}) has either an exact or approximate (with
accuracy  $O(h^{(M+1)/2})$, $M\ge 0$) solution $\Psi$ in the class of
trajectory-coherent functions. Using  an approach similar to the linear case,
we derive an infinite \HAEH system 
for the nonlinear equation (\ref{0.5}) and its finite
$M$-truncations for $ \;\{\bigl(\vec P(t,\h),\vec X(t,\h)\bigr),\;
\Delta_{\alpha\beta}(t,\hbar),\;\;$ $|\alpha|+|\beta|\leqslant
M\}$. Details are provided in Sect.~3. In particular, the
principal ($M=0$) Ehrenfest system for the nonlinear equation of
self-consistent field has the form
\begin{equation}
\begin{aligned}
&m\dot{\vec X}=\vec P,\\
&\dot{\vec P}=-\nabla_x U(\vec X)-\varkappa\nabla_x V(\vec x,\vec y)\Big|_{\vec y=\vec x=\vec X}.
\end{aligned}\label{0.7}
\end{equation}
Note that when $\varkappa=0$, this system turns to (\ref{0.4}).
Similarly to the latter, the system (\ref{0.7}) describes $\mod\h^{1/2}$ the
trajectory where the trajectory-coherent solution
$\Psi \in {\cal P}_{\h}^t$ is localized.

\item
Given a semiclassically-concentrated  solution
of equation (\ref{0.5}) with accuracy $O(\h^{(M+1)/2})$, one
can obtain a corresponding solution to the \HAEH system of order $M$
by evaluating mean values of  operators $\hat{\vec x}$, $\hat{\vec p}$,  $\widehat{\Delta}_{\alpha\beta}$
with respect to that solution.
Our goal is to show that it works the other way around as well:
quantum characteristics can be found with accuracy $O(\h^{(M+1)/2})$ from a
solution of the Ehrenfest $M$-system.

In Section 4 we study the \HAEH $2$-system for the Hartree-type
equation and in particular its solutions corresponding to the rest
point of the classical ($M=0$) system. Based on this calculation,
in Section 5 we reconstruct the  asymptotics of the energy
spectrum for the Hartree-type equation. Examples in Section 6
illustrate general results.

\end{enumerate}

The key point of the whole approach is to obtain quantum
characteristics without explicitly solving the quantum mechanical
equation. An explicit formula for the solutions is not required
anywhere in the derivation. A basic assumption of the paper is the
existence of a semiclassically concentrated solution of the
Hartree-type equation in a class $\mathcal{P}_\h^t$. This
assumption can be justified by explicit construction of formal
asymptotic solutions using  finite dimensional \HAEH systems
\cite{BTS1,BTS2,LTS1,LTS2,LTS05}.

\section{Hartree-type equation}

By the Hartree-type equation we mean the following equation
\begin{equation} \{ -i\h\pa_t +\hat \CH_{\vk}(\Psi)\}\Psi =0, \qquad\hat
\CH_{\vk}(\Psi)=\hat \CH+\vk\hat V(\Psi),\quad \Psi\in
L_2(\BR^n_x). \label{bbst1.1} \end{equation}
Here
\begin{eqnarray} & \hat {\CH} =\CH (\hat z), \label{bbst1.2}\quad
\hat V(\Psi )= \dil_{\mathbb{R}^n} d\vec y\,\Psi^*(\vec y ,t)
V(\hat z,\hat w)\Psi (\vec y ,t), \label{bbst1.3} \end{eqnarray}
where the pseudo-differential operators  $\CH (\hat z)$ and
$V(\hat z,\hat w)$ with symbols $\CH (z)$ and  $V(z,w)$
respectively are functions of non-commutative operators
\[ \hat z=( -i\h \nabla_x
, \vec x), \qquad \hat w=( -i\h\nabla_y , \vec y), \qquad \vec
x,\vec y\in \mathbb{R}^n, \] Function  $\Psi^*$ is the complex
conjugate to   $\Psi$, $\vk$ is a real parameter, $\h > 0$ is a
small parameter. The operators $\hat z$ and $\hat w$ satisfy the
following commutation relations
 \begin{eqnarray} &\left[ \hat z_k,\hat z_j \right]
=\left[ \hat w_k,\hat w_j \right] = {i}\h J_{kj},
\label{1.kom}\\ &
\left[ \hat z_k,\hat w_j \right] = 0,\qquad k,j=\overline {1, 2n},\label{2.kom} \end{eqnarray}
 where $J =\|J_{kj}\|_{2n\times 2n}$
 is the standard symplectic matrix
$$
J=\left(\begin{array}{cr}0&-\mathbb{I}\\\mathbb{I}&0\end{array}
\right)_{2n\times 2n},
$$
and $[\hat A,\hat B]=\hat A\hat B-\hat B\hat A$ denotes the commutator of $\hat A$ and $\hat B$.

In this paper all functions of non-commutative operators are Weyl-ordered
\cite{Karas2,Groot}.
The action of the  operator
 $\hat\CH$ in this case can be written as
\begin{eqnarray}
&\hat{\CH}\Psi(\vec x,t,\h) =\dac1{(2\pi
\h)^n}\int\limits_{\BR^{2n}}d\vec yd\vec p \exp\Big(\frac i\h\
\lan\vec x-\vec y, \vec p \ran\Big) \CH\Big(\vec p,\frac{\vec x+\vec
y}{2}\Big) \Psi(\vec y,t,\h),\label{bbst1.4}
\end{eqnarray}
where $\CH(z)=\CH(\vec p,\vec x)$ is the Weyl symbol of the operator
$\hat{\CH}$, and  $\lan\vec x,\vec p\ran=\sum_{k=1}^n x_k p_k$.
\\

\noindent
{\bf Remark.} In the  particular case when the Weyl symbols of operators $\CH (\hat z)$
and $V(\hat z,\hat w)$ in (\ref {bbst1.3}) have the form
$$
\CH (z)=\frac {{\vec p\,}^2}{2m}+U(\vec x),\quad V(z,w)=V(\vec x,\vec y)
$$
the equation (\ref{bbst1.1}) yields the equation of the self-consistent field
in the form
(\ref{0.5}). This differential equation with integral nonlinearity
plays a fundamental role in quantum theory and nonlinear optics
\cite{Lai,Lai90} and in the theory of the Bose-Einstein condensate
\cite{shapovalov:PITAEVSKII}. In the latter the solution $\Psi$
represents the wave function of the condensate, while the
non-local potential $V(\vec x,\vec y)$ describes the interaction
of condensate's
particles with external field.\\

In this paper we deal with asymptotic solutions of the equation
(\ref{bbst1.1}) localized in the usual mathematical sense rather
than in the sense of (\ref{0.4}): namely, functions or formal
series $\Psi(\vec x,t,\h)$ must belong to the Schwartz space with
respect to the variables $\vec x\in\BR^n$. We require the Weyl
symbols $\CH(z)$ and $V(z, w)$ of the operators   $\hat\CH$ and
$V(\hat z, \hat w)$ in (\ref {bbst1.3}) to belong to one of the
$T_+^m$ classes \cite[p.~13]{MasFed}: they must be smooth
functions of at most polynomial growth with all derivatives, such
that the following conditions hold.

\begin{assum} \label{assum1}
The functions $\CH(z)$ and $V(z,w)$ are infinitely differentiable
for all $z\in \BR^{2n}$ and $w\in \BR^{2n}$, and for any
multi-indices
$\alpha$, $\mu$ $\in\mathbb{Z}^{2n}_+$ 
there exist constants $C_\alpha$, $C_{\alpha\mu}$ and $m\ge 0$
such that
\begin{eqnarray*}
\Big| \dac{\pa^{|\alpha|}\CH(z)}{\pa z^\alpha}\Big| \le C_\alpha
(1+|z|)^m, \quad \Big| \dac{\pa^{|\alpha  + \mu |}V(z, w)} {\pa
z^\alpha  \pa w^\mu}\Big|\le C_{\alpha\mu} (1+|z|)^m (1+|w|)^m.
\end{eqnarray*}
\end{assum}
The notations here are as follows:
\begin{equation}
\label{bbst1.4q}
\begin{array}{rcl}
&\alpha=(\alpha_1,\alpha_2,\dots,\alpha_{2n}), \quad \alpha_j\ge 0, \quad
|\alpha|=\alpha_1+\alpha_2+\dots+\alpha_{2n}, \\ 
 & z^\alpha=z_1^{\alpha_1} z_2^{\alpha_2}\dots  z_{2n}^{\alpha_{2n}},
\qquad \dac {\pa^{|\alpha |}V(z)}{\pa z^\alpha}=
\frac{\pa^{|\alpha|}V(z)}{\pa z_1^{\alpha_1}\pa
z_2^{\alpha_2}\dots \pa z_{2n}^{\alpha_{2n}}}.
\end{array}
\end{equation}
Note that for our method it is essential to have smooth symbols
$\CH(z)$ and $V(z,w)$. Asymptotics for Hartee-type equations with
singularities are a  subject of a number of publications, see
e.g.\ \cite{KarasPeres01a} and references therein.

Now we introduce a vector space in which asymptotic solutions to the equation
(\ref{bbst1.1}) will be sought.

\section{Class of trajectory-coherent functions}

We will  construct  asymptotic solutions of equation (\ref{bbst1.1})
with the following features: they have a form of generalized solitary waves
and   singularly depend of the small parameter $\h\to 0$.

Such a solution relies on a phase space trajectory $Z(t,\h)=(\vec
P(t,\h),\vec X(t,\h))$ and is trajectory-coherent in the sense of
(\ref{0.14}), (\ref{xpbar}). We denote the class of
trajectory-concentrated functions by $\CP_\h^t(Z(t,\h))$ and
define it more precisely as
\begin{eqnarray}
\CP_\h^t=\CP_\h^t\big(Z(t,\h)\big) 
\displaystyle=\biggl\{\Phi :\Phi (\vec x,t,\h)=
\va\Bigl(\frac{\De\vec x}{\sqrt{\h}},t,\h\Bigr)
\exp\Bigl[{\frac{i}{\h}(S(t,\h)+ \lan\vec P(t,\h),\De\vec x \ran
)}\Bigr]\biggr\}, \label{bbst1.5}
\end{eqnarray}
where function $\va(\vec\xi ,t,\h)$ belongs to the Schwartz space
$\mathbb{S}$ with respect to variables $\vec\xi\in\mathbb{R}^n$,
is a smooth function of  $t$, and regularly depends on $\h $ as
$\h\to0$ (the term function is used throughout in the sequel,
although $\va(\vec\xi ,t,\h)$ may be in fact a formal series in
powers of $\hbar^{1/2}$). Here $\De\vec x=\vec x-\vec X(t,\h)$.
The real function $S(t,\h)$ and  $2n$-vector-function $Z(t,\h)$
also regularly depend on $\h$ as $\h\to 0$. When an asymptotic
solution of equation (\ref{bbst1.1}) is being constructed, these
functions, as well as the amplitude $\va(\vec\xi ,t,\h)$  are to
be determined. Short notation $\CP_\h^t$ for $\CP_\h^t(Z(t,\h))$
will be used when it does not lead to confusion, and $||\cdot ||$
will denote the $L_2$-norm for functions from $\CP_\h^t$. In the
expression for the norm of a function from class $\CP_\h^t$ the
argument $t$ may be omitted, and we write $\|\Phi (t)\|^2$=
$\|\Phi \|^2$.

It will be shown in Section 3 that the functions $Z(t,\hbar)$ and
$S(t,\hbar)$ are uniquely determined  by the \HAEH system
corresponding to the 
Hamiltonian of  equation (\ref{bbst1.1}).
In the linear case $(\vk=0)$ the vector-function $Z(t,0)$  and the scalar
function $S(t,0)$, defined by Hamiltonian function $\CH(\vec p,\vec x)$, are the
classical-mechanics phase space trajectory and the classical action respectively.
As an example of amplitude, dynamical coherent states for quadratic Hamiltonians
in the form of Gaussian function can be given:
\[ \va(\vec\xi,t)=\exp\Bigl[\frac i2\lan\vec\xi,Q(t)\vec\xi\ran\Bigr]f(t), \]
where $Q(t)$ is a complex symmetric matrix with positive imaginary part,
and the time dependent factor $f(t)$ is given by
\[ f(t)=\sqrt[4]{\Im\,Q(t)}\exp\Bigl[-\frac i2\inl_0^t\Im\,Q(\tau)\,d\tau\Bigr] \]
(see for details \cite{Bagre}).
\\

Consider important properties of functions from class $\CP_\h^t$.
Their proofs are given in \cite{Bagre}. We briefly reproduce some
of them  in  Appendix B.
\\

{\bf 1.}
Let $\Phi$ belong to the class $\CP_\h^t(Z(t,\h))$.
Introduce operator $\{\De{\hat z}\}^\alpha$
with Weyl symbol  $(\De z)^\alpha=(
\De z_1)^{\alpha_1}\cdot\ldots
\cdot(\De z_{2n})^{\alpha_{2n}}$,  and
\[\De z=z-Z(t,\h)=(\De\vec p,\De\vec x),\qquad \De\vec p=\vec p-\vec P(t,\h),
\qquad \De\vec x=\vec x-\vec X(t,\h) .\]
Then the
following asymptotic estimations for moments
$\De_\a(t,\h)$ of order $|\a|$, $\a\in \BZ_+^{2n}$ hold
\begin{equation}
\De_\alpha(t,\h)=\frac{\lan\Phi|\{\De{\hat z}\}^\alpha|\Phi \ran}
{\|\Phi\|^2}=O\big(\h^{|\alpha|/2}\big),\quad\h \to
0.\label{bbst1.8}
\end{equation}
\\


Denote by $\hat O(\h^\nu)$ an operator $\hat F$ such that for any function
$\Phi$, from the class $\CP_\h^t(Z(t,\h))$,
the  asymptotic estimate holds $$ \frac{\|\hat
F\Phi\|}{\|\Phi\|}=O(\h^\nu),\qquad\h\to 0.  $$
\\

{\bf 2.} The following asymptotic formula holds
\begin{equation}
\{\De{\hat z}\}^\alpha =\hat O({\h}^{|\alpha|/2}),\quad
\alpha\in\mathbb{Z}^{2n}_+,\quad\h \to 0,\label{bbst1.9}
\end{equation}
in particular
\begin{equation}
\{\De{\hat x_k}\} =\hat O({\h}^{1/2}),\quad
\{\De{\hat p_j}\} =\hat O({\h}^{1/2}),\quad k,j=\overline{1,n}.
\label{bbst1.9a}
\end{equation}

{\bf 3.} For functions $\Phi(\vec x,t,\h)\in\CP_\h^t(Z(t,\h))$
the following limits hold
\begin{eqnarray}
&&\lim _{\h \to 0} \frac 1{\| \Phi\|^2} |\Phi(\vec x,t,\h)|^2=
\de (\vec x-\vec X(t,0)),\label{bbst1.11}\\
&&\lim_{\h\to 0} \frac 1{\| \tilde\Phi\|^2} |\tilde\Phi(\vec
p,t,\h)|^2= \de (\vec p-\vec P(t,0)),\label{bbst1.12}
\end{eqnarray}
where $\tilde\Phi(\vec p,t,\h)=F_{\h,\vec x\to\vec p}\Phi(\vec
x,t,\h)$, $F_{\h,\vec x\to\vec p}$ is $\h^{-1}$-Fourier transform
 {\rm\cite{MasFed}}.

Denote by $\lan\hat A(t)\ran_{\Phi}$ the mean value of a self-adjoint in
 $L_2(\BR^n_x)$ operator $\hat A(t)$,
$t\in\BR^1$, calculated with respect to the function $\Phi(\vec
x,t,\h)\in\CP_\h^t$. Then we have

{\bf 4.} For a function $\Phi(\vec x,t,\h)\in\CP_\h^t(Z(t,\h))$
and an operator $\hat A(t,\h)$ with Weyl symbol
$A(z, t,\h)$ satisfying the first inequality in Assumption 1, the following equality holds
\begin{eqnarray}
&\displaystyle\lim _{\h \to 0} \lan\hat A(t,\h)\ran_{\Phi}=\lim _{\h \to
0} \frac 1{\|\Phi\|^2} \lan\Phi(\vec x,t,\h)|\hat
A(t,\h)|\Phi(\vec x,t,\h)\ran \vspace{14pt}=A(Z(t,0),t,0).\label{bbst1.13}
\end{eqnarray}

The limiting nature of conditions (\ref{bbst1.11}), (\ref{bbst1.12}) and
asymptotic character of estimations (\ref{bbst1.8})--(\ref{bbst1.9a}),
holding in the class of trajectory-concentrated functions ${\cal P}_{\h}^t$
allows the  construction of an approximate solution $\Psi_{\rm as}=\Psi_{\rm as}(\vec x,t,\h)$
of the Hartree-type equation for any finite time interval $[0,T]$,
in the following sense
\begin{eqnarray} & \Bigl[-i\h\dac\pa{\pa
t}+\hat\CH+\vk\hat V(\Psi_{\rm as})\Bigr] \Psi_{\rm
as}=O(\h^q),\label{zadacha1}\\ & \Psi_{\rm
as}\in\CP_\h^t(Z(t,\h),S(t,\h)), \quad t\in[0,T],\label{zadacha2}
\end{eqnarray}
where $O(\h^q)$ denotes a function $g^{(q)}(\vec
x,t,\h)$ which represents the error for equation (\ref{bbst1.1}), and
the error obeys the estimate \begin{equation} \max_{0\le t\le T}
\|g^{(q)}(\vec x,t,\h)\|=O(\h^q), \qquad \h\to0.\label{nevyazka}
\end{equation}
Following  paper \cite{Bagre} and having in mind properties
{\rm(\ref{bbst1.11}), (\ref{bbst1.12}), we call such a function
$\Psi_{\rm as}(\vec x,t,\h)$ a {\em semiclassically-concentrated
solution} ($\bmod  \h^\alpha$, $\h\to 0$) for a Hartree-type
equation (\ref{bbst1.1}).

The semiclassically concentrated ($\bmod\h^\alpha$) solution
$\Psi^{(N)}(\vec x,t,\h)$  of the Hartree-type equation
is a formal asymptotic solution evolving from
the initial state
$\Psi_0(\vec x,\h)$  chosen in the class of
trajectory-concentrated functions $\CP^0_\h(z_0,S_0)$.
Here $z_0=(\vec p_0,\vec x_0)$ is an arbitrary point of the
phase space
$\BR^{2n}_{px}$, and the constant
$S_0=S(0,\h)$ can be set equal to zero without loss of generality
 We will denote the class of initial $(t=0)$ trajectory-concentrated functions by
$\CP^0_\h(z_0)$. Explicitly,
\begin{eqnarray}
\CP_\h^0\big(z_0\big) =\biggl\{\psi :\psi (\vec x,\h)=
\va_0\Bigl(\frac{\vec x-\vec x_0} {\sqrt\h},\h\Bigr)
\exp\Bigl\{\frac i\h\lan\vec p_0,
\vec x-\vec x_0\ran\Bigr\},\quad
\va_0(\vec\xi,\h)\in\BS(\BR^n_\xi)\biggr\}, \label{vid}
\end{eqnarray}

Let us give two important examples of the amplitude function in the
initial state (\ref{vid}).

First, $$
\va_0(\vec\xi)=e^{-\lan\vec\xi,A\vec\xi\ran/2}, 
$$
where the $n\times n$-matrix $A$ is real symmetric and positive
definite.
In this case relation
(\ref{vid}) describes the  {\em Gaussian wave packet}.

Second,
$$ \va_0(\vec\xi)=e^{i\lan\vec\xi,Q\vec\xi\ran/2}H_\nu(\Im\,Q\vec\xi),
$$ where the $n\times n$-matrix $Q$ is complex symmetric and has
positive definite imaginary part
$\Im\,Q$, and
$\nu=(\nu_1,\dots,\nu_n)$
is a multi-index of
the multi-dimensional Hermite polynomial
$H_\nu(\vec\eta)$, $\vec\eta\in\BR^n$
\cite{Beitman2}.
In this case $\psi\in {\cal P}_{\h}^0(z_0)$ (\ref{vid}) defines
the {\em Fock state of a multi-dimensional oscillator}.

A  construction of a semiclassically-concentrated solution ($\bmod\h^\alpha$), of the problem  (\ref{bbst1.1}) with initial state from  (\ref{vid})
is based on the solution of the \HAEH system,
to which we turn our attention now.

\section{The \HAEH system of equations}

Let symbols $\CH(z)$  $V(z,w)$ satisfy  Assumption 1. Then operator
$\CH (\hat z)$ (\ref{bbst1.2})
is self-adjoint with respect to inner product
$\lan\Psi|\Phi\ran$ in space  $L_2(\BR_x^n)$,
and operator  $V(\hat z,\hat w )$ (\ref{bbst1.3}) is self-adjoint
for the inner product in space $L_2(\BR^{2n}_{xy})$.
Thus the norm of the exact solutions of
(\ref{bbst1.1}) is preserved by time evolution:
$ \|\Psi(t)\| =\|\Psi_0\|$.
The mean value $\lan\hat A\ran=\lan\hat A\ran_\Psi=\lan \Psi|\hat A|\Psi\ran
$ of an operator
$\hat A(t)=A(\hat z,t)$,
calculated with respect to these solutions obeys
\begin{eqnarray}
& \dac{d}{dt}\lan\hat A(t)\ran =\Bigl\lan\frac{\pa\hat A(t)} {\pa
t}\Bigr\ran +\frac{i}{\h}\lan[{\CH}(\hat z),\hat A (t)]\ran 
+\dac{i\vk}{\h}\Bigl\lan\!\int\! d\vec y\,\Psi^*(\vec y,t,\h)
  [V(\hat z,\hat w),\hat A (\hat z, t)]\Psi(\vec y,t,\h)\!\Bigl\ran\!,
\label{bbst2.1}
\end{eqnarray}
as an implication of the Heisenberg equation for evolution of operators.
Equation (\ref{bbst2.1})
is called the {\em Ehrenfest equation for operator $\hat A(t)$
and
function}\ $\Psi(\vec x,t,\h)$.
Our choice of this terminology is justified by analogy with  the linear
case ($\vk=0$) in which equation (\ref{bbst1.1}) becomes the
Schr\"odinger equation, while equation (\ref{bbst2.1}) is the Ehrenfest
equation \cite{Ehrenfest}.

To derive the \HAEH system from the Ehrenfest equation (\ref{bbst2.1}) we
take for $\Psi$ a solution of the Hartree-type equation
(\ref{bbst1.1})
in the class of trajectory-concentrated functions, and
for $\hat A$ operators
$\hat z =(\hat {\vec p},\hat {\vec x})$ and
$\{\Delta\hat z\}^\alpha$ with Weyl symbols
$\{\Delta z\}^\alpha$, $\alpha\in\BZ_+^{2n}$, where
\begin{equation}
\Delta z = z -Z(t,\h),  \quad
Z(t,\h)=\lan \Psi(t)|\hat z|\Psi(t)\ran.
  \label{bbst2.2}
\end{equation}
We represent operators $\CH (\hat z,t)$ and $V(\hat z, \hat w,t)$ in the series form
\begin{eqnarray}
&&\CH (\hat z)=\CH(Z(t,\hbar))+\dum_{|\mu|=1}^\infty \dac1{\mu!}\CH_\mu(Z(t,\hbar))\{\Delta\hat z\}^\mu,
\label{bbst2.2ch}\\
&&V(\hat z,\hat w)=V(Z(t,\hbar),Z(t,\hbar))+
\dum _{|\nu|=1}^\infty \sum_{|\mu|=2}^\infty \dac1{\nu!\mu!}V_{\mu\nu}(Z(t,\hbar))
\Delta_\mu \{\Delta\hat z\}^\nu,\cr
&&\quad \CH_\mu(z,t)=\dac{\pa^{|\mu|}\CH ( z,t)}{\pa z^\mu},\quad
V_{\mu\nu}(z,t)= \frac{\pa^{|\mu+\nu|} V( z, w,t)}
 {\pa z^\mu \pa w^\nu}\Big|_{\omega=z},
\quad \mu,\nu\in\BZ_+^{2n},
 \nonumber
\end{eqnarray}
where $\Delta_\mu$ as defined in (\ref{bbst1.8}). Thus to derive
the system we need to evaluate commutators $[\hat z_k,
\{\Delta\hat z\}^\mu]$ and $[\{\Delta\hat z\}^\nu,\{\Delta\hat
z\}^\mu]$ for $k=\overline{1,2n}$ and $|\mu|\ge 1$, $|\nu|\ge 1$.
This has been done in the linear case ($\vk=0$) \cite{Bagre,
69,88a,88}) using the formula for composition of Weyl symbols
$A(z)$ and $B(z)$ (see e.g. Appendix in \cite{Karas2}),  defining
the symbol $C(z)$ of the product of operators
 $\hat C=$ $\hat A\,\hat B$
\begin{equation}
C(z)=A\Big(\stackrel{2}{z}+\frac{i\h}2J\frac{\stackrel{1}{\pa}}
{\pa z}\Big) B(z)=B \Big(\stackrel{2}{z}-\dac{i\h}2J
\frac{\stackrel{1}{\pa}}{\pa z}\Big) A( z).
\end{equation}
Here the number over an operator refers to the order of its action
onto the target function.

This way we obtain an infinite system of equations for $Z(t,\hbar), \Delta_\alpha(t,\hbar)$.
Keeping only the moments up to order $N$,
we obtain the following finite system of equations
\begin{eqnarray}
&&\displaystyle\dot z=\sum_{|\mu|=0}^N \frac 1{\mu!}
J\Big(\CH_{z\mu}(z)\De_\mu +\tvk\sum_{|\nu|=0}^N\frac 1{\nu!}
 V_{z\mu\nu}(z)\De_{\mu}\De_{\nu}\Big),\cr
&&\dot\De_{\alpha}=\dum_{|\mu+\gamma|=0}^N
\Big(-i\h\Big)^{|\gamma|-1}\,
\frac{[(-1)^{|\gamma_p|}-(-1)^{|\gamma_x|}]\alpha!\beta!
\theta(\alpha-\gamma)\theta(\beta-\gamma)}
{\gamma!(\alpha-\gamma)!(\beta-\gamma)!\mu!}\times \label{bbst2.4}\\
&&\quad\times\displaystyle \Big(\CH_\mu(z)+\tvk\sum_{|\nu|=0}^N\frac
1{\nu!}
 V_{\mu\nu}(z)\De_{\nu}\Big)\De_{\alpha-\gamma +J\beta-J\gamma}-
\sum_{k=1}^{2n}\dot Z_k \alpha_k \De_{\alpha(k)}\nonumber
\end{eqnarray}
with initial conditions
\begin{eqnarray}
& z\big|_{t=0}=z_0=\lan\psi|\hat z|\psi\ran, \qquad
\De_\alpha\big|_{t=0}=\lan\psi|\{\hat z-z_0\}^\alpha|\psi\ran,
\label{nul} \qquad
\alpha\in\mathbb{Z}^{2n}_+, \quad |\alpha|\le N.
\end{eqnarray}
Here $\tvk=\vk\|\psi(\vec x,\h)\|^2$, and  $\psi(\vec
x,\h)$ is the initial state from $\CP_\h^0(z_0)$ (\ref{vid}),
\begin{eqnarray}
& \CH_{z\mu}(z)=\dac{\pa^{|\mu|}\CH _z( z)}{\pa z^\mu},\quad
V_{z\mu\nu}(z)= \frac{\pa^{|\mu+\nu|} V_z( z, w)}
 {\pa z^\mu \pa w^\nu}\Big|_{\omega=z},
 \qquad \theta(\alpha-\beta)=\prod\limits_{k=1}^{2n}\theta(\alpha_k-\beta_k),
 \label{bbst2.5}\\[2pt]
&\alpha=(\alpha_p,\alpha_x), \quad J\alpha=(\alpha_x,\alpha_p) ,
\quad
\alpha(k)=(\alpha_1-\de_{k1}, \dots,
\alpha_{2n}-\de_{k2n}).\nonumber
\end{eqnarray}

As in the linear case
($\vk=0$) (see \cite{Bagre})
the system (\ref{bbst2.4}) will be called
{\em \HAEH system of order $N$}.
Due to estimates (\ref{bbst1.8})
this system is equivalent in the class of trajectory-concentrated states 
to  the Hartree-type equation
(\ref{bbst1.1}) with precision
$O(\h^{(N+1)/2})$.

Introduce notations
\begin{equation}
\begin{array}{l}
\!\!\FH(z,w)=\CH(z)+\tvk V(z,w),\quad
 \FH_{z}(z)=\FH_{z}(z,w)\Big|_{w=z}\!=\bigg\|\dac{\pa\FH(z,w)}{\pa z_j}\bigg|_{w=z}\bigg\|_{1\times2n}\!,\\
 \FH_{zz}(z)=\FH_{zz}(z,w)\Big|_{w=z}=\bigg\|\dac{\pa^2\FH(z,w)}{\pa z_j\pa z_k}\bigg|_{w=z}\bigg\|_{2n\times2n}\!.
 \end{array}\label{bbst2.6d}
\end{equation}

Then for $N=0$  the \HAEH system (\ref{bbst2.4}) has the form
\begin{equation}
 \displaystyle\dot z=J\FH_z(z), \label{bbst2.6_0}
\end{equation}
and for $N=2$ we obtain
\begin{equation}
\left\{
\begin{array}{l}
 \displaystyle\dot z=J\pa_z\Bigl[1+\frac 12\lan\pa_z,
 \De_2\pa_z\ran+\frac 12\lan\pa_w,
 \De_2\pa_w\ran\Bigr]\FH(z,w)\Big|_{w=z},\\[8pt]
 \dot\De_2=J\FH_{zz}(z)\De_2-\De_2 \FH_{zz}(z)J,
 \quad\De_2^\intercal=\De_2,\end{array}\right.\label{bbst2.6}
\end{equation}
where
 $\De_2$ is a symmetric $(2n\times 2n)$-matrix of the second moments
($\De_2=\|\De_{ji}\|_{2n\times 2n}$)
\begin{equation}
\label{shapovalov:1.12} \Delta_2(t)=\left (
\begin{array}{cc}
\sigma_{pp}(t)& \sigma_{px}(t)\\
\sigma_{xp}(t) &\sigma_{xx}(t)
\end{array} \right ),
\end{equation}
with $(n\times n)$-blocks
\begin{eqnarray*}
&&\sigma_{pp}(t)=\|\sigma_{p_k p_l}(t)\|_{n\times n}=\|\langle
\Delta\hat p_k \Delta\hat p_l\rangle)\|_{n\times n}, \quad 
\sigma_{xx}(t)=\|\sigma_{x_k x_l}(t)\|_{n\times n}=\|\langle
\Delta x_k
\Delta x_l\rangle\|_{n\times n}, \\[2 pt]
&&\sigma_{xp}(t)=\|\sigma_{x_k p_l}(t)\|_{n\times n}=
\Big\|\displaystyle\frac{1}{2}\langle \Delta x_{k} \Delta \hat
p_l+\Delta\hat p_l \Delta x_k \rangle\Big\|_{n\times n};
\end{eqnarray*}

System (\ref{bbst2.6}) can be written in an equivalent form
introducing matrix $A(t)$ via relation
\[\De_2(t)= A(t)\De_2(0) A^+(t). \]
Then the system becomes
\begin{equation}
\left\{
\begin{array}{l}\dot z = J\pa_z\Bigl[1+\dac 12\lan\pa_z,
 A\De^0_2A^+\pa_z\ran+\frac 12\lan\pa_w,
 A\De^0_2A^+\pa_w\ran\Bigl]\FH(z,w)\Big|_{w=z},\\[8pt]
 \dot A= J \FH_{zz}(z) A, \qquad A(0)=\mathbb{I}.
\end{array}\right.\label{ks23}
\end{equation}

Note that the initial state function can be excluded from the initial
conditions (\ref{nul}) for the system (\ref{bbst2.6}) if  they satisfy an infinite system
of inequalities
\footnote{For $|\alpha+\beta|\leqslant 4$ the system of inequalities is
well known (see e.g. \cite{DoMa2},
and the bibliography there).}
(generalized Heisenberg uncertainty conditions)
and for $t=0$ satisfy estimations (\ref{bbst1.8}).
Obviously, all Heisenberg inequalities are consistent with
equations (\ref{bbst2.4}).

The uncertainty relation corresponding to the \HAEH system 
(\ref{bbst2.6}) can be rewritten as the condition that
matrix $\De_2(t)+\dac{i\h}2J$ is positive definite \cite{29} (see also \cite{DoMa2,Bagre}).

The fact that equation
(\ref{bbst2.4})
is a finite system of ordinary
differential equations for functions
 $z$, $\De_\alpha$  equivalent to problem
(\ref{bbst1.1})
 with precision $O(\h^{(N+1)/2})$, suggests that there is a mechanical
 system with finite degrees of freedom described by
(\ref{bbst2.4}),
and thus an approximate semiclassical model of a
Hartree-type equation allows an exact interpretation in frame of classical
mechanics.
The number of degrees of freedom of such a mechanical system grows with
precision $N$.
A study of such classical systems with quantum origin  by methods of
classical mechanics constitutes a separate interesting direction of
research \cite{Sadov96, KondraSad04}.

\section
{The \HAEH system: \\ the small dispersions approximation}

Denote by  $\hat \fg$ operators
\begin{equation}
\hat\fg=\bigg({\hat z}_{j},\hat\Delta_2^{kl}=\dac12(
 \De\hat {z}_{k}\De\hat {z}_{l}+\De\hat {z}_{l}\De\hat {z}_{k});\quad j,k,l=\overline{1,2n}, k\le l\bigg).
 \label{kvs7q}
\end{equation}
Their mean values in state  $\Psi (\vec x,t)$ with initial
condition $\Psi (\vec x,t)\Big|_{t=0}=\psi(\vec x)$, are denoted
by
\begin{equation}
\fg_{\psi}(t,\h)=\lan\Psi (\vec x,t)|\hat\fg|\Psi (\vec x,t)\ran=\bigg(Z(t,\h,\fg^0_{\psi}),
 \De_2(t,\h,\fg^0_{\psi})\bigg).\label{kvs7w}
\end{equation}
The mean values $\fg_\psi(t,\h)$ obey \HAEH system
(\ref{bbst2.6})
with initial condition  $\fg_{\psi}(0,\h)=\fg^0_{\psi}=\lan\psi|\hat \fg|\psi\ran$.
When it does not lead to  confusion the explicit dependence of the initial function will be omitted in the notations.

Estimations (\ref{bbst1.8})  suggest that the \HAEH system
can be solved approximatly with respect to a small parameter $\h$
using the expansions
\begin{equation}
\fg(t,\h)=\fg^{(0)}(t,\h)+\h \fg^{(1)}(t,\h) +\ldots,\label{kvs7ww}
\end{equation}
or
\begin{equation} \begin{array}{l}Z(t)=Z^{(0)}(t)+\h Z^{(1)}(t) +
\ldots,\\[6pt] \De_2(t) = \De^{(0)}_2(t,\h) +\h  \De^{(1)}_2(t,\h)+ \ldots.\end{array}\label{kvs7}
\end{equation}
Here $\fg^{(0)}(t,\h)=\Big(Z^{(0)}(t), \De^{(0)}_2(t,\h)\Big)$ is the principal term of the solution of system  (\ref{bbst2.6}).
Substituting (\ref{kvs7ww}) into system (\ref{bbst2.6})
 we obtain equations for
$z^{(0)}=Z^{(0)}$, $z^{(1)}=Z^{(1)}$, and
$\De_2^{(0)}$     with precision
$O(\h^{3/2})$
\begin{equation}
\left\{
\begin{array}{l}{\dot z}^{(0)} = J \FH_z(z^{(0)}), \\[8pt]  {\dot z}^{(1)} =
J\FH_{zz}(z^{(0)})z^{(1)} + F(z^{(0)},\De^{(0)}_2),\\[8pt]
\dot\De_2^{(0)} = J\FH_{zz}(z^{(0)})\De_2^{(0)}-\De_2^{(0)}\FH_{zz}(z^{(0)})J.
\end{array}\right.\label{kvs8}
\end{equation}
Here
\begin{equation}
F(z,\De_2)=\dac{1}{2\h} J\pa_{z}\Sp\Bigl\{\Bigl[\FH_{zz}(z,w)+ \tvk
V_{ww}(z,w)\Bigr]\De_2\Bigr\}\Big|_{w=z}.\label{kvs11a}
\end{equation}
The first equation of system  (\ref{kvs8}) coincides with (\ref{bbst2.6_0}) and
is similar to the classical-mechanical Hamilton system in the linear case, however,
in the nonlinear case ($\tvk\ne 0$) the system is not Hamiltonian.

Consider the following auxiliary system of equations
which can be referred as the {\it pseudo-system-in-variations}
for the solution $Z^{(0)}(t)$
\begin{equation}
\dot a_k = J\FH_{zz}(Z^{(0)}(t))a_k \quad k=\overline{1,n}, \label{kvs9}
\end{equation}
with normalization condition
\begin{equation}
\{a_k(t),a_l(t)\}=\{a_k^*(t),a_l^*(t)\}=0, \quad
\{a_k^*(t),a_l(t)\}=-2 i\de_{kl},\label{kvs10}
\end{equation}
where $\{v,u\}$ is a skew symmetric inner product in $\BR^{2n}$
\begin{eqnarray}
&\{v,u\}=\lan v, J^\intercal u\ran = \vec W_a^\intercal\,\vec Y_b -
\vec Y_a^\intercal \,\vec W_b,\label{kvs11}\\
& v=\left(\begin{array}{c}\vec W_a\\ \vec Y_a\end{array}\right),
\quad u=\left(\begin{array}{c}\vec W_b\\ \vec
Y_b\end{array}\right). \nonumber
\end{eqnarray}

If a solution $z^{(0)}=Z^{(0)}(t)$  as well as
a complete set of solutions $a_k(t)$ of the pseudo-system-in-variation are known
then the general solution of the two last equations in
(\ref{kvs8})  has the form
\begin{eqnarray}
&& Z^{(1)}(t)=\dum_{k=1}^n \left[ b_k(t)a_k(t)+b_k^*(t)a_k^*(t)
\right],\label{kvs12}\\
&& \De_2^{(0)}(t) = A(t)\CD A^t(t),\label{kvs12a}
\end{eqnarray}
where scalar functions $b_k(t)$ and the $2n\times 2n$-matrix $A(t)$ are as follows
\begin{eqnarray}
&& b_k(t)=- \dac1{2 i}\dil_0^t\{a_k^*(t),\tilde F(t)\} dt + b_k,\quad
\tilde F(t)=F(Z^{(0)}(t),\De_2^{(0)}(t))
,\label{kvs12aq}\\[6pt] &&
A(t)=\Big(a_1(t),a_2(t),\ldots,a_n(t),a_1^*(t),
a_2^*(t),\ldots,a_n^*(t)\Big).\label{kvs12aw}
\end{eqnarray}
Here $b_k$ are constants of integration,
$\CD$ is an arbitrary $2n\times 2n$ constant matrix and  $F(Z^{(0)}(t),\De_2^{(0)}(t))$
is defined in (\ref{kvs11a}). Thus in this approximation
the total solution is determined by solutions of the
modified classical system
(\ref{bbst2.6_0}) and pseudo-system-in-variations (\ref{kvs9}).

\section
{Energy spectrum and the \HAEH system for quantum means.}

Consider the  problem of defining the energy spectrum for a Hartree-type Hamiltonian $\hat \CH_{\vk}$
from the dynamical \HAEH system for quantum means. Denote by
$S[\fg_\psi(t,\h)]$ the generalized action  \cite{BTS1,BTS2} along
the trajectory  $\fg_\psi(t,\h)$  (\ref{kvs7w}) of the \HAEH
system {\rm (\ref{bbst2.6})}
\begin{eqnarray}
&&\!\!\!\!\!\!\!\!S\Big[\fg_\psi(t,\h)\Big]=\!\dil^t_0\!dt\Bigl[\Big\lan\vec
P(t,\fg^0_\psi),\dot{\vec X}(t,\fg^0_\psi)\Big\ran 
-\FH\Big(z,w\Big)-\!\frac\tvk2\!\Sp\!\!\Big(\!
V_{ww}(z,w)\De_2(t,\fg^0_\psi)\!\Big)\!\Bigr]\!\Big|_{\!w=z=Z(t,\fg^0_\psi)}\!\!-\cr
&&- \h\Big\lan P^{(0)}(t,\fg^0_\psi),X^{(1)}(t,\fg^0_\psi)\Big\ran+
\h\lan p_0,X^{(1)}_0\ran.\label{kvs3a}
\end{eqnarray}
Substituting (\ref{kvs7}) in (\ref{kvs3a}) we get
\begin{eqnarray}
&&\!\!\!\!\!\!\!\!S\Big[\fg_\psi(t,\h)\Big]=\!\dil^t_0\!dt\Bigl[\Big\lan\vec
P^{(0)}(t,\fg^0_\psi),\dot{\vec X}^{(0)}(t,\fg^0_\psi)\Big\ran -\FH\Big(z,w\Big)-\cr &&
- \h\tvk
\lan V_w(z,w),Z^{(1)}(t,\fg^0_\psi)\ran -
\!\frac\tvk2\!\Sp\!\!\Big(\!
V_{ww}(z,w)\De_2^{(0)}(t,\fg^0_\psi)\!\Big)\!\Bigr]\!\Big|_{\!w=z=Z^{(0)}(t,\fg^0_\psi)}
+O(\h^{3/2}).\label{kvs3aa}
\end{eqnarray}

The connection between the energy  spectrum and the time-dependent
solution $\fg_{\psi}(t,\h)$
is established as follows

\begin{demo}
Let the Hartree-type stationary equation
\begin{equation}
\widehat\CH_\vk(\va_\nu)\va_\nu = E_\nu\va_\nu,\qquad \nu\in\BZ_+^n\label{kvs1}
\end{equation}
have  pure discrete non-degenerate spectrum, and functions $\va_\nu$ form a basis
in $\CP^0_\h(z_0)$ {\rm (\ref{vid})}. Let a solution $\fg(t,\h)$
of system {\rm (\ref{bbst2.6})} exist in the class of
quasi-periodic functions {\rm \cite{levitan}:}
\begin{equation}
\fg(t,\h) = \sum_{|\nu+\nu'|=0}^\iy \fg_{\nu\nu'}(\h)
e^{i\omega_{\nu \nu'}t}, \label{kvs2}
\end{equation}
and satisfy the generalized Heisenberg inequalities.
Here  $\fg_{\nu\nu'}(\h)$
are generalized Fourier coefficients for function
$\fg (t,\h)$.
Then the energy levels $E_\nu$ and  
the frequencies $\omega_{\nu \nu'}$
 are connected by the following relation $(\!\bmod\,\h^{3/2})$
\begin{equation}
\h\omega_{\nu\nu'}=E_\nu-E_{\nu'}+\dac1tS[\fg_{\va_\nu}(t,\h)]-
\dac1tS[\fg_{\va_{\nu'}}(t,\h)],
\qquad \nu\nu'\in\BZ^n_+.\label{kvs3}
\end{equation}
\end{demo}

{\bf Proof}.
At the initial moment of time function $\fg(t,\h)$ satisfies
the generalized Heisenberg inequalities. Thus there exists a function
$\psi\in {\cal P}_{\h}^0(z_0)$ such that with the accuracy $O(\h^{3/2})$,
$\fg(0,\h)=\fg_\psi(0,\h)=\lan\psi|\hat \fg|\psi\ran$ (see \cite{Bagre}).

Since vectors $\va_\nu$ form a complete set and belong to $\CP^0_\h(z_0)$,
every semiclassically-concentrated solution
of (\ref{bbst1.1})  with the same Hamiltonian as in
(\ref{kvs1}),
and initial state $\psi\in\CP^0_\h(z_0)$,
can be represented in the form
(see Appendix A):
\begin{equation}
\Psi(\vec
x,t)=\exp\Big(\frac{i}{\h}S[\fg_\psi(t,\h)]\Big)\sum_{|\nu|=0}^\iy
C_{\nu}\exp\Bigl(
-\frac{i}{\h}E_{\nu}t-\frac{i}{\h}S[\fg_{\va_\nu}(t,\h)]\Bigr)\va_{\nu}(\vec
x) +O(\h^{3/2}).\label{kvs4}
\end{equation}

Using  function $\Psi(\vec x, t)$ in this form for
evaluation of the mean value of operators $\hat\fg$ (\ref{kvs7q}) and comparing with (\ref{kvs2}) we get
(\ref{kvs3}), which completes the proof.  \bigskip

If the ground level of energy
$E_0$   as well as the complete set of 
frequencies
$\omega_{\nu\nu'}$ are given,  then the problem of reconstruction of the
entire  spectrum $E_\nu$  is a well known problem in spectroscopy.

The invariant manifolds of classical Hamiltonian systems are known
to be important for classification of spectral series in the
linear theory ($\vk=0$) \cite{BeD2}.

We are able to solve the problem in the semiclassical approximation
using invariant manifolds
$\fg^{(0)}(t)$ of system (\ref{kvs8}) for classification of the  spectral
series.

In the simplest case when the invariant set consists of a single rest point
$z^{(0)}=Z^{(0)}(t)=\const$ 
it follows from (\ref{kvs8}) that
\begin{equation}
\FH_z(z^{(0)})=\pa_{z}\Bigl[\CH(z) +\tvk V(z,w)\Bigr]\Big|_{w=z=z^{(0)}}=0
.\label{kvs13}
\end{equation}
Identify the rest point $Z^{(0)}$
 with point $z_0$ which defines class $\CP^0_\h(z_0)$ (\ref{vid}).
In other words we take initial states localized at the rest point $z_0=Z^{(0)}$.

We will proceed under the following assumption.

\begin{assum}
Let the symplectic $(2n\times 2n)$-matrix $J\FH_{zz}(z_0)$ evaluated at the rest point $z_0$
 have $n$ distinct pure imaginary eigenvalues
$i\Omega_k$, $\Omega_k>0$, $k=\overline{1,n}$ (and $n$ eigenvalues complex conjugate to them, $-i\Omega_k$, $k=\overline{1,n}$).
\end{assum}

In the linear theory, Assumption 2 implies stability of the rest
point in the linear approximation \cite{Maslov2}.

Note that under Assumption 2,
 solutions of the pseudo-system-in-variation (\ref{kvs9}) have the form
\begin{equation}
a_k(t) = \exp (i\Omega_kt)f_k,\quad k=\overline{1,n},\label{kvs14}
\end{equation}
where
 $f_k$ is the eigenvector of the pseudo-matrix-in-variations, evaluated at the rest point
\begin{equation}
J\FH_{zz}(z_0)f_k = i\Omega_k f_k, \qquad
\quad\Omega_k\ne\Omega_j,\quad j,k=\overline{1,n}.\label{kvs15}
\end{equation}
The eigenvectors $f_k$, $k=\overline{1,n}$ are normalized, without
loss of generality, by condition (\ref{kvs10}).

\begin{demo} \label{st5-2}
Under Assumption 2, the energy spectrum of the Hartree-type
operator {\rm (\ref{kvs1})} can be found as
\begin{equation}
E_\nu =\FH(z_0,z_0)
+\h\sum_{k=1}^n\widetilde\Omega_k \bigg(\nu_k +
\dac12\bigg)+O(\hbar^{3/2}),\quad \label{kvs22}
\end{equation}
where multi-index $\nu=(\nu_1,\dots,\nu_n)$ has all non-negative components,
$\FH(z,w)$ is defined in  {\rm (\ref{bbst2.6d})} and
\begin{eqnarray}
&&\widetilde\Omega_k=\Omega_k+\dac\tvk2 \lan
f^*_k,V_{ww}(z_0,z_0)f_k\ran+\cr && \quad +\Re\dum_{j=1}^n
\frac{\tvk}{2\Omega_j} \lan V_w(z_0,z_0),f_k\ran \lan
f^*_k,\pa_z\ran \lan f^*_j, 
[\FH_{zz}(z,w)+\tvk V_{ww}(z,w)]
f_j\ran\Big|_{z=w=z_0}.\label{kvs18a}
\end{eqnarray}
\end{demo}
For $\nu=0$ our result (\ref{kvs22})  coincides with results obtained in \cite{Simenog}
with precision  $O(\hbar^{3/2})$.

 To prove Statement  \ref{st5-2}
 we need several Lemmas.

\begin{lem} Under Assumption 2, the spectrum of Hartree-type operator
{\rm (\ref{kvs1})} is equidistant ({\rm mod}$\,\hbar^{3/2}$) and
can be found as
\begin{eqnarray}
&& E_{\nu} = E_{0}+ \h\sum_{k=1}^n \widetilde\Omega_k\nu_k,\label{kvs18r}
\end{eqnarray}
where $E_0=${\rm const} is the ground energy level and $\widetilde \Omega_k$ are defined in {\rm(\ref{kvs18a})}.
\end{lem}

\noindent
{\bf Proof}.
As the first step, we establish the linear relation between $\Omega_k$,
 $k=\overline{1,n}$ and
$\omega_{\nu\nu'}$ in the Statement 5.1.
Using (\ref{kvs14}), formulas (\ref{kvs12}) and (\ref{kvs12a})
with (\ref{kvs11a}),  (\ref{kvs12aq}), (\ref{kvs12aw}) become
\begin{equation}
\begin{array}{l}
Z^{(1)}(t)=\dum_{k=1}^n \Re\left[\{f_k^*, \FF_k(t)\}f_k +b_k
e^{i\Omega_kt}f_k
\right],\qquad b_k=\const,\\[9pt]
\De_2^{(0)}(t)=\dum_{j,l=1}^n \frac12\left(f_jf_l^+ +f_j^*f_l^\intercal
\right)\CD_{jl}e^{i(\Omega_j-\Omega_l)t}, \quad f_l^+=(f_l^*)^\intercal,\quad\CD_{jl}=\const.
\end{array} \label{kvs16}
\end{equation}
Here the $2n$-vector $\FF_k(t)$ is defined by the relation
\begin{equation}
\FF_k(t)=\dac{1}{2\h} J\pa_{z}\Sp\Bigl\{\Bigl[\FH_{zz}(z,w)+ \tvk
V_{ww}(z,w)\Bigr]\CF_k(t)\Bigr\}\Big|_{w=z=z_0},\label{kvs11aq}
\end{equation}
where  the $(2n\times2n)$-matrix $\CF_k(t)$ has the following structure
\begin{equation}
\CF_k(t)=\dum_{j,l=1}^n \frac1{2(\Omega_k+\Omega_j-\Omega_l)}\left(f_jf_l^+ +f_j^*f_l^\intercal
\right)\CD_{jl}e^{i(\Omega_j-\Omega_l)t}.\label{kvs11aw}
\end{equation}

Using the estimate (\ref{bbst1.8}), solution of the \HAEH system
(\ref{bbst2.4}) can be written in the form
\begin{equation}
\fg(t,\h) = \bigg(z_0+\h Z^{(1)}(t),\De_2^{(0)}(t)\bigg)+O(\h^{3/2}). \label{kvs2q}
\end{equation}
Introduce vectors
\begin{eqnarray}
&\vec\mu=(\mu_1,\ldots,\mu_n),\quad\vec\nu=(\nu_1,\ldots,\nu_n),\quad\vec\Omega=(\Omega_1,\ldots,\Omega_n), \label{kvs2qo}
\end{eqnarray}
and let $|\vec \mu|$ denote the sum of the absolute values of
vector's components. Substitute (\ref{kvs16}) into (\ref{kvs2q})
and rewrite it as
\begin{eqnarray}
&\fg(t,\h) = \dum_{|\vec\mu|\le 2} \fg_{\mu}(\h)
e^{i\lan\vec\mu,\vec\Omega\ran t} +O(\h^{3/2}). \label{kvs2qr}
\end{eqnarray}
The right hand side in (\ref{kvs2qr}) is an almost periodic
function with $n$ frequencies $\Omega_1,...\Omega_n$. The
exponents in (\ref{kvs2qr}) are linear combinations of the
frequencies $\Omega_k$, $k=\overline{1,n}$ with integer
coefficients $\mu_k$, and thus  linearly depend on the multi-index
$\mu\in\BZ^n_+$. Consequently, frequencies  $\omega_{\nu\nu'}$
(\ref{kvs2}), must be linear functions of $\nu$ and  $\nu'$.
Taking into account that by interchanging
 $\nu$ and  $\nu'$ frequencies  $\omega_{\nu\nu'}$ (\ref{kvs3}) change
 sign, and by preceding arguments, we get
\begin{eqnarray}
&\omega_{\nu\nu'}=\lan\vec\nu-\vec\nu',\vec\Omega\ran
=\lan \vec\mu , \vec\Omega\ran, \quad |\vec \mu|\le 2. \label{kvs2qw}
\end{eqnarray}
\\

The step 2 of our proof consists in recalculation of the right hand side
of the formula (\ref{kvs3}).

Energy level $E_\nu$  with precision $O(\h^{3/2})$ can be found from the mean value of $\hat \CH_\vk$ in the stationary state $\va_\nu\in
{\cal P}_{\h}^0(z_0)$ as follows
\begin{eqnarray}
&&E_\nu=\lan\va_\nu|\hat\CH_\vk(\va_\nu)|\va_\nu\ran=\FH(z_0,z_0)+ \h\tvk \lan
V_w(z_0,z_0),Z^{(1)}_{\nu}\ran +\nonumber\\[4pt] &&\qquad
+\dac{1}{2}\Sp\left\{\Big[
\FH_{zz}(z_0)+\tvk V_{ww}(z_0,z_0)\Big]\De_2^{\nu}\right\}
+O(\h^{3/2}).\label{kvs19}
\end{eqnarray}
Here the equality (\ref{kvs13}) was used.

Note that for the stationary state $\va_\nu$ we have for coefficients in (\ref{kvs16})
$b_k^{\nu}=0$, $\CD_{jl}^\nu=\CD_{l}^{(\nu)}\de_{jl}$ since the functions are time-independent
\begin{equation}
Z^{(1)}_{\nu}=\dum_{k=1}^n
\Re[{\{f_k^*, \FF^{\nu}_k\}}f_k ],\qquad
\De_2^{\nu}=\dum_{l=1}^n \frac12\left(f_lf_l^+ +f_l^*f_l^\intercal
\right)\CD_l^{(\nu)}.
\label{kvs16a}
\end{equation}
Here vector  $\FF^{\nu}_k$ is defined  by  (\ref{kvs11aq}) with
\begin{equation}
\CF_k(t)=\CF^{\nu}_k=\dum_{j=1}^n
\frac1{2\Omega_k}{\CD_{j}^{(\nu)}}\left(f_jf_j^+
+f_j^*f_j^\intercal \right)\!.\label{kvs11aqw}
\end{equation}
Substituting
 formulas (\ref{kvs16a}) into (\ref{kvs19}) we get after some calculations
 using (\ref{kvs15})
\begin{eqnarray}
&E_\nu=\FH(z_0,z_0)+ \h\tvk \lan
V_w(z_0,z_0),Z_1^{(\nu)}\ran 
+\dum_{k=1}^n \Bigl[\Omega_k+\dac\tvk2 \lan
f^*_k,V_{ww}(z_0,z_0)f_k\ran\Bigr]
\CD_k^{(\nu)}+O(\h^{3/2}).\label{kvs5}
\end{eqnarray}

Similarly, from (\ref{kvs3aa}) we find
\begin{eqnarray}
&&\dac1tS[\fg_{\va_\nu}]=-\FH(z_0,z_0)- \h\tvk
\lan V_w(z_0,z_0),Z_1^{(\nu)}\ran 
-\tvk\dum_{k=1}^n
\lan f^*_k,V_{ww}(z_0,z_0)f_k\ran \CD_k^{(\nu)},\label{kvs3b}
\end{eqnarray}
Now substitute  (\ref{kvs5}),(\ref{kvs3b}) in (\ref{kvs3}) to find that
$$
\hbar \omega_{\nu\nu'}=\sum_{j=1}^n \Omega_j (D_j^{(\nu)}-D_j^{(\nu')}).
$$
From this equation along with (\ref{kvs2qw}) we can find $D_j^{(\nu)}$
in terms of $D_j^{(0)}$ as follows
\begin{equation}
\CD_j^{(\nu)}=\CD_j^{(0)}+\hbar \nu_j,\quad j=\overline{1,n}.\label{kvs21a}
\end{equation}
Constants $\CD_j^{(0)}$ will be defined later.

To illustrate how the equation (\ref{kvs21a}) was obtained we
temporarily consider case $n=2$, without loss of generality. 
Then condition $|\vec\mu|\le 2$ in (\ref{kvs2qw}) reduces our consideration
effectively to the three cases.

Case 1. If $\mu_1=1$, $\mu_2=0$ then $\omega_{\nu\nu'}=\Omega _1$ and
$\nu_1-\nu'_1=1$, $\nu_2-\nu'_2=0$. Consequently,
$$
\hbar \Omega_1=\Omega_1 (D_1^{(\nu_1,\nu_2)}-D_1^{(\nu_1-1,\nu_2)})+
\Omega_2 (D_2^{(\nu_1,\nu_2)}-D_2^{(\nu_1-1,\nu_2)}),
$$
and thus
$$
D_1^{(\nu_1,\nu_2)}=D_1^{(\nu_1-1,\nu_2)}+\hbar,\quad
D_2^{(\nu_1,\nu_2)}=D_2^{(\nu_1-1,\nu_2)}.
$$
Case 2. If $\mu_1=0$, $\mu_2=1$ then $\omega_{\nu\nu'}=\Omega _2$ and
$\nu_1-\nu'_1=0$, $\nu_2-\nu'_2=1$. Consequently,
$$
\hbar \Omega_2=\Omega_1 (D_1^{(\nu_1,\nu_2)}-D_1^{(\nu_1,\nu_2-1)})+
\Omega_2 (D_2^{(\nu_1,\nu_2)}-D_2^{(\nu_1,\nu_2-1)}),
$$
and thus
$$
D_1^{(\nu_1,\nu_2)}=D_1^{(\nu_1,\nu_2-1)},\quad
D_2^{(\nu_1,\nu_2)}=D_2^{(\nu_1,\nu_2-1)}+\hbar.
$$
Case 3. If $\mu_1=1$, $\mu_2=1$ then $\omega_{\nu\nu'}=\Omega_1+\Omega _2$ and
$\nu_1-\nu'_1=1$, $\nu_2-\nu'_2=1$. Consequently,
$$
\hbar (\Omega_1+\Omega_2)=\Omega_1 (D_1^{(\nu_1,\nu_2)}-D_1^{(\nu_1-1,\nu_2-1)})+
\Omega_2 (D_2^{(\nu_1,\nu_2)}-D_2^{(\nu_1-1,\nu_2-1)}),
$$
and thus
$$
D_1^{(\nu_1,\nu_2)}=D_1^{(\nu_1-1,\nu_2-1)}+\hbar,\quad
D_2^{(\nu_1,\nu_2)}=D_2^{(\nu_1-1,\nu_2-1)}+\hbar.
$$
Note that the third case result can be decomposed into the first two.
It is  apparent  that similar derivations are possible in
any dimension $n\ge 2$, and that  equation (\ref{kvs21a}) holds.
\\

As the third step of the proof, we finally  obtain the statement of Lemma 5.1.
Using (\ref{kvs16a}) for $Z^{(1)}_{\nu}$ with (\ref{kvs11aqw}) for $\FF^\nu_k$, 
from (\ref{kvs11a}) and (\ref{kvs16}),  we find
\begin{eqnarray}
&&\lan V_w(z_0,z_0),Z_1^{(\nu)}\ran =\Re\dum_{k=1}^n
\dum_{j=1}^n\frac{1}{2\Omega_j} \lan V_w(z,w),f_k\ran\times\cr &&
\quad \times\lan f^*_k,\pa_z\ran \Bigr\lan f^*_j,
\Bigl[\FH_{zz}(z,w) +\tvk V_{ww}(z,w)\Bigr]\Big|_{z=w=z_0}
f_j\Bigr\ran\CD_j^{(\nu)}.\label{kvs55}
\end{eqnarray}
From (\ref{kvs5}), using (\ref{kvs21a})  and (\ref{kvs55}) we get
\begin{eqnarray}
&& E_{\nu} = \FH(z_0,z_0)+
\sum_{j=1}^n \widetilde\Omega_j\Big[\CD_j^{(0)}+
\hbar\nu_j\Big],\label{kvs18}
\end{eqnarray}
where $\widetilde\Omega_j$ is given by (\ref{kvs18a}).
Finally, denote by
\begin{equation}
E_0=\FH(z_0,z_0)+ \sum_{j=1}^n
\widetilde\Omega_j\CD_j^{(0)}\label{kvs20}
\end{equation}
to obtain (\ref{kvs18r}). Thus the lemma is proved.

\begin{lem}
Under Assumption  2, the ground energy level for the Hartree-type operator
{\rm (\ref{kvs1})}
is given  ({\rm mod} $\,\hbar^{3/2}$) by
\begin{eqnarray}
&& E_0=\FH(z_0,z_0)+ \dac\h2\sum_{j=1}^n
\widetilde\Omega_j.\label{kvs18w}
\end{eqnarray}
\end{lem}
\noindent {\bf Proof}. The ground energy level  $E_0$ is given by
(\ref{kvs20}), which follows from (\ref{kvs18}) for $\nu=0$.
Constants $\CD_k^{(0)}$ will be chosen in order to minimize the
uncertainty condition and obey the quantization condition
(\ref{kvs3}). This idea is supported by physical models including
the harmonic oscillator and the Coulomb potential \cite{Bargm,
Faris} (see \cite{DoMa2}  and references there).

As we have mentioned above, matrix $\De_2(t)+\dac{i\h}2J$ is positive definite.
Thus for any vector $v$ we have
$$
v^+\bigg[\De_2(t)+\dac{i\h}2J\bigg]v\geqslant 0,
$$
where equality corresponds to minimization of the uncertainty relation.

Take  $v = J  f_j, j =
\overline{1,n}$. Then using (\ref{kvs16a}) with $\nu=0$ and orthogonal condition (\ref{kvs10})
(which is valid for vectors $f_k$, as well as for $a_k(t)$) we have
\[
\CD_j^{(0)} \geqslant \dac\h2.
\]
So, choosing
\begin{equation}
\CD_j^{(0)}=\dac\h2,\label{kvs21b}
\end{equation}
we minimize the uncertainty relation, and obtain (\ref{kvs18w}).

Then formula (\ref{kvs22}) directly follows from
(\ref{kvs18r})  and (\ref{kvs18w}).
Thus statement \ref{st5-2} is proved.

\section{Spectrum for the oscillator with  nonlinear Gaussian
potential }
\subsection{Spectrum for the oscillator in constant magnetic field and
nonlinear Gaussian potential. }

In this section we illustrate the method described above with
an example of a Hartree-type equation (\ref{bbst1.1}) whose linear part corresponds to
an oscillator in a constant magnetic field, while the nonlinear part
is described via Gaussian potential.
The linear part
$\hat {\mathcal H}$ has the form
\begin{equation}
\label{shapovalov:0.5} \hat {\mathcal H}
=\displaystyle\frac{1}{2m}(\hat{\vec p}-\frac{e}{c}\vec A(\vec
x))^2 +\displaystyle\frac{k}{2}\vec x^2.
\end{equation}
The external field in the operator (\ref{shapovalov:0.5})
is a superposition of a constant magnetic field $\vec H=(0,0,H)$ with
vector potential $\vec A=\displaystyle\frac{1}{2}\vec H\times
\vec x$ and an oscillatory field with scalar potential
$\displaystyle\frac{k}{2}\vec x ^2$.
The non-local operator
$\hat V(\Psi )$ in (\ref{bbst1.3}) has the form
\begin{equation}
\label{shapovalov:0.6}
\hat V(\Psi )=\int_{\mathbb{R}^3}V(\vec x,\vec y)|\Psi(\vec y,t)|^2 d\vec
y,\qquad
V(\vec x, \vec y)=V_0\exp\left
[-\displaystyle\frac{(\vec x-\vec y)^2}{2\gamma^2} \right ].
\end{equation}
Here $H,V_0$, $k$, $\gamma$,  $e$, $c$ are real parameters of
the model.

We will be using notations $\omega_{H}$, $\omega_0$, and $\omega_{\rm nl}$ for
the cyclotron frequency, oscillator frequency, and nonlinear frequency
respectively
\begin{equation}
\omega_{H}=\displaystyle\frac{e H}{mc},\qquad
 \omega_0= \sqrt{\displaystyle\frac{k}{m}},\qquad
 \omega_{\rm nl}=\sqrt{\displaystyle\frac{|\tilde\varkappa
V_0|}{m\gamma^2}},
\label{shapovalov:0.6a}
\end{equation}
where
$
\tilde\varkappa=\varkappa\|\Psi \|^2.
$
We also introduce
\begin{equation}
\omega_a=\omega_0\sqrt{1+\displaystyle
\left(\frac{\omega_{H}}{2\omega_0}\right)^2}.
\label{shapovalov:0.6c}
\end{equation}

To construct a solution $\Psi\in$ $\mathcal P_\hbar^t(Z(t,\hbar))$ of equation
(\ref{bbst1.1}), (\ref{shapovalov:0.5})--(\ref{shapovalov:0.6}),
we are using a phase space trajectory $Z(t,\hbar)=(\vec P(t,\hbar),\vec X(t,\hbar))$
which obeys (\ref{0.14}), (\ref{xpbar}).

Using notations (\ref{bbst2.6d}) we have
\begin{eqnarray}
&&{\mathcal H}(z)=\displaystyle\frac{1}{2m}\vec P^2+
\displaystyle\frac{m\omega_a^2}{2}(X_1^2+ X_2^2)+
\displaystyle\frac{m\omega_0^2}{2}X_3^2+
\frac{\omega_{H}}{2}(P_1 X_2-P_2 X_1); \label{shapovalov:1.14}\\
&&\label{shapovalov:1.15}
\FH_{z}(z)=
\mathcal{H}_z(z)=
 \left ( \!\!\!\begin{array}{c}
\displaystyle\frac{1}{m}P_1+\frac{\omega_{H}}{2}X_2 \\[10 pt]
\displaystyle\frac{1}{m}P_2-\frac{\omega_{H}}{2}X_1\\[10 pt]
\displaystyle\frac{1}{m}P_3\\[10 pt]
\displaystyle-\frac{\omega_{H}}{2}P_2+m\omega_a^2X_1 \\[10 pt]
\displaystyle\frac{\omega_{H}}{2}P_1+m\omega_a^2X_2 \\[10 pt]
m\omega_0^2X_3
\end{array} \!\!\!\right ),
\end{eqnarray}
where $z=(\vec P,\vec X)$, $\vec P=(P_1,P_2,P_3)$, $\vec X=(X_1,X_2,X_3)$, 
and notations
 (\ref{shapovalov:0.6a}),
(\ref{shapovalov:0.6c}) are used. The matrix of the second derivatives becomes
\begin{eqnarray}
\label{shapovalov:1.16} &&\FH_{zz}(z)=\left (
\begin{array}{cc}
\FH_{pp}(z)& \FH_{px}(z)\\
\FH_{xp}(z) &\FH_{xx}(z)
\end{array} \right ),
\\[6 pt] &&
\label{shapovalov:1.17} \FH_{pp}(z)=\|\FH_{p_k
p_l}(z)\|_{3\times3}=\mbox{diag}\Big(\displaystyle\frac{1}{m},\displaystyle\frac
{1}{m}, \displaystyle\frac{1}{m}\Big);
\\[6 pt]
&&\label{shapovalov:1.18} \FH_{xx}(z)=\|\FH_{x_k
x_l}(z)\|_{3\times3} 
=\mbox{diag}\big(m(\omega_a^2-\eta\omega_{\rm nl}^2 )
,m(\omega_a^2-\eta\omega_{\rm nl}^2), m(\omega_0^2-
\eta\omega_{\rm nl}^2) \big);\\[6 pt]
&& \label{shapovalov:1.19}
\FH_{px}(z)=\|\FH_{p_kx_l}(z)\|_{3\times3}=\left (
\begin{array}{ccc}
0& \displaystyle\frac{\omega_{H}}{2}&0 \\[6 pt]
-\displaystyle\frac{\omega_{H}}{2}&0&0 \\[6 pt]
0&0&0
\end{array}\right ).
\end{eqnarray}
Here $\eta={\rm sign}(\tilde\varkappa V_0)$.

Recall, that to find the spectrum corresponding to the Hamiltonian in
(\ref{bbst1.1}) we need only bounded solutions of the
\HAEH system. 
The first equation of system
(\ref{kvs8})  describes
${Z_0}(t,\hbar)$, and can be integrated independently from the other
equations of the system.
 The last equation of
(\ref{kvs8}) describes the second moments ${\Delta}_2(t)$ and
depends on the solution of the first one.
Therefore we start with solving the first equation of (\ref{kvs8}).
The simplest stationary solution of it is the zero solution
\begin{eqnarray} &&Z_0(\hbar)=\Big(\vec P_0(\hbar),\vec
X_0(\hbar)\Big)^\intercal =(0,0,0,0,0,0)^\intercal. \label{shapovalov:2.3}
\end{eqnarray}

Then the corresponding eigenvalue problem (\ref{kvs15}) has solutions
$\Omega_1=\omega_+$, $\Omega_2=\omega_-$, $\Omega_3=\omega_s$, where the  Ritz frequencies
$$
\omega_{+}=\displaystyle\sqrt{\omega_a^2-\eta\omega_{\rm
nl}^2} + \displaystyle\frac{\omega_{H}}{2},\quad
\omega_{-}=\displaystyle\sqrt{\omega_a^2-\eta\omega_{\rm nl}^2}-
\displaystyle\frac{\omega_{H}}{2},\quad
\omega_{\rm s}=\sqrt{\omega_0^2-\eta\omega_{\rm nl}^2},
$$
and the eigenvectors are
\begin{eqnarray}
&&
f_1=\displaystyle\frac{1}{\sqrt 2}(g_0,ig_0,0,-
\frac{i}{g_0},\frac{1}{g_0},0)^\intercal, \nonumber\\
&&f_2=\displaystyle\frac{1}{\sqrt 2}(g_0,-ig_0,0,-
\frac{i}{g_0},-\frac{1}{g_0},0)^\intercal, \label{shapovalov:2.7}\\
&&f_3=\displaystyle
(0,0,g_{\rm s},0,0, -\frac{i}{g_{\rm
s}})^\intercal.\nonumber
\end{eqnarray}
Here $g_0=\sqrt{\displaystyle\frac{m}{2}(\omega_{+}+\omega_{-})}$,
$g_{\rm s}=\sqrt{m\omega_{\rm s}}$.
The solutions $a_j(t)$ of (\ref{kvs9}) are found by (\ref{kvs14})
and are normalized  by condition (\ref{kvs10}).
They form a matrix
$A(t)$ (\ref{kvs12aw}) which we rewrite in the block form
\begin{equation}
\label{shapovalov:2.8} A(t)=\left (
\begin{array}{cc}
B(t)& B^*(t)\\
C(t)& C^*(t)
\end{array} \right ),
\end{equation}
where matrices $B(t)$, $C(t)$  have the following form
\begin{eqnarray*}
&\label{shapovalov:2.8a} B(t)=\left (
\begin{array}{ccc}
\displaystyle\frac{g_0 e^{i\omega_{+}t}}{\sqrt 2}&
\displaystyle\frac{g_0 e^{i\omega_{-}t}}{\sqrt 2}&0\\[10 pt]
\displaystyle\frac{ig_0 e^{i\omega_{+}t}}{\sqrt 2}&
\displaystyle\frac{-i g_0e^{i\omega_{-}t}}{\sqrt 2}&0\\[10 pt]
0&0 &g_{\rm s} e^{i\omega_{s}t}
\end{array} \right ), 
\quad C(t)=\left (
\begin{array}{ccc}
\displaystyle\frac{-i e^{i\omega_{+}t}}{\sqrt{2}g_0}&
\displaystyle\frac{-i e^{i\omega_{-}t}}{\sqrt{2}g_0}&0\\[10 pt]
\displaystyle\frac{e^{i\omega_{+}t}}{\sqrt{2}g_0}&
\displaystyle\frac{-e^{i\omega_{-}t}}{\sqrt{2}g_0}&0\\[10 pt]
0&0 &\displaystyle\frac{-ie^{i\omega_{s}t}}{g_{\rm s}}
\end{array} \right ).
\end{eqnarray*}
Using (\ref{kvs21a}), (\ref{kvs21b}), and (\ref{kvs12a}) we find a solution   of
the last equation of system  (\ref{kvs8}) in the block form (\ref{shapovalov:1.12}) in terms of the blocks
$B(t)$,$C(t)$ of the matrix  $A(t)$ (\ref{shapovalov:2.8}) as follows
\begin{eqnarray*}
&\sigma_{xx}(t)=\displaystyle\frac{\hbar}{2}\Big(C(t)D(\nu)
C^+(t)+C^*(t)D(\nu)
C^\intercal(t)\Big),\\
&\sigma_{pp}(t)=\displaystyle\frac{\hbar}{2}\Big(B(t)D(\nu)
B^+(t)+B^*(t)D(\nu)
B^\intercal(t)\Big),\\
&\sigma_{px}(t)=\displaystyle\frac{\hbar}{2}\Big(B(t)D(\nu)
C^+(t)+B^*(t)D(\nu) C^\intercal(t)\Big),
\end{eqnarray*}
where the diagonal matrix
$D(\nu)=\mbox{diag}(\nu_1+1/2,\nu_2+1/2,\nu_3+1/2)$,
$\nu_1,\nu_2,\nu_3=\overline{0,\iy}$, and  the symbol $(^*)$
denotes the Hermitian conjugate matrix.
Matrices $\sigma_{xx}$, $\sigma_{pp}$ are diagonal and their
explicit form is as follows
\begin{eqnarray}
&&\sigma_{xx}(t)=\displaystyle\frac{\hbar}{m}\mbox{diag}\Big(
\displaystyle\frac{\nu_1+\nu_2+1}{\omega_{+}+\omega_{-}},
\displaystyle\frac{\nu_1+\nu_2+1}{\omega_{+}+\omega_{-}},
\displaystyle\frac{2\nu_3+1}{2\omega_{s}}\Big) \cr
&&\sigma_{pp}(t)=\displaystyle\frac{\hbar m}{4}\mbox{diag}\Big(
(\omega_{+}+\omega_{-})(\nu_1+\nu_2+1),\label{shapovalov:2.9}
(\omega_{+}+\omega_{-})(\nu_1+\nu_2+1),
 2\omega_{s}(2\nu_3+1)\Big). \nonumber
\end{eqnarray}
The non-zero elements of the matrix $\sigma_{xp}(t)$ are
$\sigma_{p_1x_2}(t)=$ $-\sigma_{p_2x_1}=$ $\hbar(\nu_1-\nu_2)/2$.

After substitution (\ref{shapovalov:2.7}) into (\ref{kvs22}),
(\ref{kvs18a}) and taking into account that
for $V$ defined by (\ref{shapovalov:0.6}),
the vector $V_w(z_0,z_0)=0$,
and the only non-zero elements of the $6\times 6$ matrix $V_{ww}(z_0,z_0)$ are $(V_{ww})_{jj}=-V_0/\gamma^2$
for $j=4,5,6$,
we obtain the energy spectrum ${E}_\nu$ of the Hamiltonian $\hat\CH_{\vk}$
(\ref{bbst1.1}), (\ref{shapovalov:0.5})--(\ref{shapovalov:0.6})
\begin{eqnarray}
\label{shapovalov:3.6} &&{E}_\nu=\displaystyle \tilde\varkappa V_0
+\hbar\left [\Big(\omega_{+}-\displaystyle\frac{\eta\omega_{\rm
nl}^2}{\omega_{+}+\omega_{-}}\Big)\Big(\nu_1+\frac 12\Big)+
\right.\nonumber\\
&&\left.\quad+\Big(\!\omega_{-}-\displaystyle\frac{\eta\omega_{\rm
nl}^2}{\omega_{+}+ \omega_{-}}\!\Big)\Big(\!\nu_2+\frac
12\!\Big)\!+ \!\Big(\!\omega_{s}-
\displaystyle\frac{\eta\omega_{\rm nl}^2} {2\omega_{\rm s}
}\!\Big)\Big(\!\nu_3+\frac 12\!\Big)\!\right ] +O(\hbar^{3/2}).
\end{eqnarray}
Note that in the case of zero magnetic field $H=0$, a
similar expression for
spectrum was obtained in \cite{LTS2}.

\subsection{One-dimensional case}

Consider equation (\ref{bbst1.1}) with linear operator $\hat {\mathcal
H}(t)$ in the form
\begin{equation}
\label{shapovalov:4.1}
\hat {\mathcal H}
=\displaystyle\frac{\hat{ p}^2}{2m} +\displaystyle\frac{k}{2}
x^2,
\end{equation}
and the nonlinear operator $\hat V(\Psi (t))$ as follows
\begin{equation}
\label{shapovalov:4.2} \hat V(\Psi (t))\Psi(
x,t)=\int_{-\infty}^{+\infty}V(x, y)|\Psi(y,t)|^2 d y \Psi( x,t),
\quad V(x, y)=V_0\exp\left [-\displaystyle\frac{(x-y)^2}{2\gamma^2}
\right].
\end{equation}
In the absence of a magnetic field ($H=0$), the cyclotron frequency
(\ref{shapovalov:0.6a}) is equal to zero ($\omega_{H}=0$), and thus from
(\ref{shapovalov:0.6c}) we find $\omega_a=\omega_0$.
Then for the Ritz frequencies we have
\[
\omega_{+}=\omega_{-}=\omega_{\rm
s}=\sqrt{\omega_{0}^2-\eta\omega_{\rm nl}^2}.
\]
The \HAEH system with accuracy
$O(\hbar^{3/2})$  for operators
(\ref{shapovalov:4.2})  has the form
\begin{eqnarray}
&&\displaystyle\left\{\begin{array}{ll}
\dot{p}=-kx,\\
\displaystyle\dot{x} =\frac{p}{m},
\end{array}\right. \label{shapovalov:4.3}\\
&&\displaystyle\left\{\begin{array}{l}
\displaystyle\dot{\sigma}_{xx}=\frac 2m\sigma_{xp}, \vspace{2mm}\\
\displaystyle\dot{\sigma}_{xp} = \frac
1m\sigma_{pp}-m\omega_{\rm s}^2 \sigma_{xx}, \\[6pt]
\displaystyle\dot{\sigma}_{pp} =-2m\omega_{\rm
s}^2\sigma_{xp}.\end{array}\right.\label{shapovalov:4.3a}
\end{eqnarray}
As a stationary solution of
subsystem (\ref{shapovalov:4.3})  we take the zero solution
\begin{eqnarray}
&&Z_0(\hbar)=\Big(P_0(\hbar), X_0(\hbar)\Big)^\intercal
=(0,0)^\intercal, \label{shapovalov:2.3a}
\end{eqnarray}

In this case matrix $A(t)$ (\ref{shapovalov:2.8})
of the pseudo-system-in-variations (\ref{kvs9}) is a $2\times 2$-matrix whose
scalar blocks $B(t)$ and $C(t)$ satisfy  equations
\begin{equation}
\dot B=-m\omega_{\rm s}^2C,\quad\displaystyle\dot
C=\frac{B}{m}.\label{shapovalov:4.4}
\end{equation}
Floquet solutions  (\ref{shapovalov:2.7}) $a(t)= \Big(B(t),C(t)\Big)^\intercal$
of the system in variations (\ref{shapovalov:2.7}) normalized by condition
$\langle a,J^\intercal a^*\rangle=2i $, $
J=\left(\begin{array}{cc}0&-1\\
1&0\end{array}\right)$ can be written in the form
\begin{eqnarray*}
a(t)= \frac{\exp(i\omega_{\rm s} t)}{\sqrt{m\omega_{\rm s}}}
\left(\begin{array}{c}
im\omega_{\rm s}\\
1\end{array}\right).
\end{eqnarray*}
Then for the energy spectrum  ${E}_n$ of the Hartree-type equation
(\ref{bbst1.1}),
(\ref{shapovalov:4.1}),
(\ref{shapovalov:4.2})
we obtain
\begin{eqnarray} &&{E}_n=\tilde\varkappa V_0
+\hbar\Big(\omega_{\rm s}- \frac{\eta\omega_{\rm nl}^2} {2\omega_{\rm
s}}\Big) \Big(n+\frac12\Big)+O(\hbar^{3/2}).\label{shapovalov:4.7}
\end{eqnarray}

\section*{Concluding remarks}

An approach to the problem of correspondence between classical and
quantum models in the nonlinear case has significant differences
from the one feasible in linear quantum mechanics. In the linear
quantum case, a   transition from quantum to classical system in
the sense of Ehrenfest requires a certain property of a
quantum-mechanical solution, namely the function has to be
trajectory-coherent (\ref{0.14}). A state which does not obey this
condition is considered to be essentially quantum, but  one which
obeys it is near-classical. For the near-classical states the
classical dynamics obtained in the limit $\hbar\to 0$ is defined
by the classical Hamilton function, and appears to be the same
regardless whether the quantum solution is localized (at each
moment of time) at a point, on a curve or on a surface. (Exact
meaning of  localization on a curve or  surface is explained e.g.
in \cite{1,MasFed}.)

For the Hartree-type equation the situation is different. Classical equations (\ref{0.7}) (or (\ref{bbst2.6_0}))
are valid only for states concentrated at a point in each moment of time. These classical equations are distinct from those
obtained in \cite{Mas1,Mas3,Mas2,Maslov01}. The latter are integro-differential equations which
describe dynamics of
the $n$-dimensional manifolds in the $2n$ dimensional phase space. It was shown that for Hartee-type
equations an implementation of
Born's approach leads to those integro-differential equations for
characteristics of a non-local
(Vlasov) equation which describes the evolution of  the classical density matrix.
(Recall that in the linear case, the  classical density matrix
obeys the local Liouville equation, whose characteristics are
trajectories of classical mechanics.) Thus dynamics of point-wise
and elongated objects have different equations in the case of
Hartree-type models. A rigorous derivation of classical equations
describing dynamics of $k$-dimensional objects ($0<k<n$) in
$2n$-dimensional phase space constitutes a separate open problem.

The \HAEH systems (\ref{bbst2.4}) are subjects of mathematical interest,
independent of their quantum origin.
Questions similar to those in the linear case, such as about their Poisson
structure, and the stability of the solutions
(including stability with respect to the nonlinearity parameter $\vk$) can
be addressed in  future study.

In this paper we have shown how the energy spectrum for the Hartree-type equation can be retrieved from
a rest-point solution of related \HAEH system. Similarly,
other quantum characteristics such as
quasi-energy spectrum, geometric and adiabatic phases, can be reconstructed from solutions of corresponding \HAEH systems.
We will attempt to show that in details in our future publications.
Note that in our approach the quantum characteristics can be found without solving the quantum equation.
This is  particularly valuable and advantageous  due to lack of general methods for solving Hartree-type equations.
\\

{\bf Acknowledgment}

The authors are grateful to our colleagues Sergey Yu. Sadov, Thomas A.
Osborn and Frank H. Molzahn for critically reading the manuscript and a number of
suggestions.

This research was supported by grant
to M.F.K. from
the Natural Sciences and Engineering Research Council of Canada.
V.V.B. and
A.Yu.T. acknowledges support from
Russian Foundation for Basic Research, grant No 050122002, and
from the President of the Russian Federation grant NSh-1743.2003.2
respectively.

\section*{Appendix A}
\def\theequation{{\rm A}.\arabic{equation}}
\setcounter{equation}{0}
\def\thesection{\rm A.}
\setcounter{samp}{0} \setcounter{teo}{0}\setcounter{demo}{0}

\begin{demo} A solution of equation {\rm(\ref{bbst1.1})} with the same Hamiltonian  as in {\rm
(\ref{kvs1})} and initial condition  $\Psi(\vec x,t)|_{t=0}= \psi\in\CP^0_\h(z_0)$ {\rm(\ref{vid})}
may be written in the form  {\rm(\ref{kvs4})},
where $\{\va_\nu(\vec x)\}^\iy_{|\nu|=0}$, $\va_\nu\in \CP^0_\h(z_0)$ is a complete set of stationary trajectory-concentrated states
of equation  {\rm (\ref{kvs1})}, with corresponding eigenvalues $E_{\nu}$, and
$S[\fg_\psi(t,\h)]$ is defined in {\rm(\ref{kvs3aa})}
\end{demo}

This statement follows from the {\it nonlinear  semiclassical  superposition principle}, which we derive first.

\begin{lem}
 Let $\Big\{\Psi_\nu(\vec x,t)\Big\}^N_{|\nu|=0}$ be a set
of semiclassically-concentrated mod $ \h^{3/2}$ solutions of equation
 {\rm
(\ref{bbst1.1})} with initial conditions
$\Psi_\nu(x,0)=\psi_\nu(\vec x)\in\CP^0_\h(z_0)$. Then function
\begin{equation}
\Psi(\vec
x,t)=\exp\Big(\frac{i}{\h}S[\fg_\psi(t,\h)]\Big)\sum_{|\nu|=0}^N
C_{\nu}\exp\Big(-\frac{i}{\h}S[\fg_{\psi_\nu}(t,\h)]\Big)\Psi_\nu(\vec
x,t) +O(\h^{3/2}),\label{kvs4rr}
\end{equation}
also is a semiclassically-concentrated mod $ \h^{3/2}$ solution of equation {\rm (\ref{bbst1.1})} with initial condition
$\Psi(\vec x,0)=\psi(\vec x)=\dum_{|\nu|=0}^N
C_{\nu}\psi_\nu(\vec x)\in\CP^0_\h(z_0)$. Here notation {\rm(\ref{kvs3aa})} was used. 

\end{lem}

\noindent {\bf Proof}. Consider associated mod $\h^{3/2}$ to
(\ref{bbst1.1}) the linearized in ${\cal P}_{\h}^t$ Schr\"odinger
equation \cite{BTS1,BTS2,LTS1,LTS05}
\begin{eqnarray}
&&\bigg(-i\hbar \frac \partial{\partial t} +\wh \FH_0(\fg_\psi(t,\h))\bigg)\Psi=0,\label{kvs4rqr}\\
&&\wh \FH_0(\fg_\psi(t,\h))=\Bigl\{\FH(z,w)+
\dac\tvk2\Sp[V_{ww}(z,w)\De_2]+\lan\FH_{z}(z),\Delta \hat
z\ran+\cr&&\quad+ \dac 12\lan  \Delta\hat z,\FH_{zz}(z)\Delta\hat
z\ran\Bigr\}\Big|_{w=z=Z (t,\h)}, \quad \Delta\hat z=\hat z-Z
(t,\h). \label{kvs4rqt1}
\end{eqnarray}
Here $\fg_\psi(t,\h)=(Z (t,\h), \Delta_2 )$ is defined in
(\ref{kvs7w}), vector  $\FH_{z}(z)$ and matrix $\FH_{zz}(z)$ are
defined in  (\ref{bbst2.6d}). Substitute for the argument of  $\wh
\FH_0(\fg_\psi(t,\h))$ (\ref{kvs4rqt1}) the expansion
$\fg(t,\h)=\fg^{(0)}(t,\h)+\h \fg^{(1)}(t,\h) $ (\ref{kvs7ww}),
where $Z^{(0)}(t)$ (\ref{kvs7}) is the principal term of the phase
space trajectory. Then with precision $O(\h^{3/2})$, we obtain
\begin{eqnarray}
&
\wh \FH_0(\fg_\psi(t,\h))={\FH}^{(2)}_\vk\Big(\fg_\psi(t,\h)\Big)+\Big\lan\FH_{z}\Big(Z^{(0)}(t)\Big),\Delta \hat z_0\Big\ran+
\dac 12\Big\lan  \Delta\hat z_0,\FH_{zz}\Big(Z^{(0)}(t)\Big)\Delta\hat z_0\Big\ran,\label{kvs4rqt}
\end{eqnarray}
where $\Delta\hat z_0=\hat z-Z^{(0)}(t)$, and ${\FH}^{(2)}_\vk\Big(\fg_\psi(t,\h)\Big)$
is defined by the following equation
\begin{eqnarray}
&&\!\!\!\!\!\!{\FH}^{(2)}_\vk\Big(\fg_\psi(t,\h)\Big)=
\Bigl\{\FH(z,w)+\h\lan\FH_{w}(z),Z^{(1)}(t)\ran+
\dac\tvk2\Sp[V_{ww}(z,w)\De_2^{(0)}(t)]\Bigr\}\Big|_{w=z=Z^{(0)}(t)}.\label{kvs4rqw}
\end{eqnarray}
A solution mod $\h^{3/2}$  of the equation (\ref{kvs4rqr}) may be written in the form 
\begin{equation}
\Psi\Big(\vec
x,t;\fg_\psi(t,\h)\Big)=\exp\Big(\frac{i}{\h}S[\fg_\psi(t,\h)]\Big)\chi\Big(\vec
x,t,Z^{(0)}(t)\Big),\label{kvs4rq}
\end{equation}
 and the equation for
$\chi\Big(\vec x,t,Z^{(0)}(t)\Big)$, taking into account (\ref{kvs3aa}), becomes
\begin{eqnarray}
&&\bigg(-i\hbar \dac d{d t} +\lan\dot{\vec P}^{(0)}(t),\Delta x_0\ran+
\dac 12\Big\lan  \Delta\hat z_0,\FH_{zz}\Big(Z^{(0)}(t)\Big)\Delta\hat z_0\Big\ran\bigg)\chi=0,
\label{kvs4rqu}\\
&&\quad \frac d{d t}=\frac\partial{\partial t}+\lan\dot{\vec X}^{(0)}(t),\nabla\ran,\quad
\Delta x_0=x-\vec X^{(0)}(t).\nonumber
\end{eqnarray}
Note that  equation (\ref{kvs4rqu}) is defined only by trajectory
$Z^{(0)}(t)$, it is linear, and it describes evolution of any
initial state from class $\CP_\h^0(z_0)$ with $z_0=Z^{(0)}(0)$.
Thus if the initial state is represented as a linear combination
\begin{equation}
\chi(\vec x,0,z_0)= \dum_{|\nu|=0}^N C_{\nu}
 \chi_\nu(\vec x,0,z_0),
\quad \chi_\nu(\vec x,0,z_0)\in\CP_\h^0(z_0) \label{kvs4rp}
\end{equation}
then
\begin{equation}
\chi\Big(\vec x,t,Z^{(0)}(t)\Big)=\sum_{|\nu|=0}^N
C_{\nu}\chi_\nu\Big(\vec x,t,Z^{(0)}(t)\Big) +O(\h^{3/2}).\label{kvs4ro}
\end{equation}

Now, from (\ref{kvs4rq}) we have $\Psi\Big(\vec x,0;\fg_\psi^0\Big)=\chi(\vec
x,0,z_0),$ and thus (\ref{kvs4rp}) can be written as
\begin{equation}
\Psi(\vec x,0)=
\dum_{|\nu|=0}^N C_{\nu}
 \Psi_\nu(\vec x,0),\quad \Psi_\nu(\vec x,0)\in \CP_\h^0(z_0).
\end{equation}

To complete the proof of the lemma it remains to observe that
taking into account  (\ref{kvs4rq}), equation (\ref{kvs4rr}) is
equivalent to (\ref{kvs4ro}) .
\\

To justify Statement A.1 we
take $N=\iy$  and $$\Psi_\nu(\vec
x,t)=\exp\Big[ -\frac{i}{\h}E_{\nu}t\Big]\va_\nu(\vec
x,\h).$$ Then from (\ref{kvs4rr}) we obtain (\ref{kvs4}).
\\

\section*{Appendix B}
\def\theequation{{\rm B}.\arabic{equation}}
\setcounter{equation}{0}
\def\thesection{\rm B.}
\setcounter{samp}{0} \setcounter{teo}{0}\setcounter{demo}{0}

}

Here we derive properties of the trajectory-coherent functions, listed in Section 2.
\\

{\bf 1.} Proof of (\ref{bbst1.8}).
Rewrite the Weyl symbol of the operator $\{\De\hat z\}^\alpha$
in the form
\[(\De z)^\alpha=(\De\vec p)^{\alpha_p}(\De\vec x)^{\alpha_x}, \quad
(\alpha_p,\alpha_x)=\alpha.\]
 Thus, in accordance with
 (\ref{bbst1.4}), we obtain the following formula for the mean value
$\sigma_\alpha(t,\h)$ of the operator $\{\De\hat z\}^\alpha$
\begin{eqnarray*}
\sigma_\alpha(t,\h) &=& \big\lan \Phi |\{\De{\hat z}\}^\alpha |
\Phi \big\ran=\frac 1 {{(2\pi\h)}^n} \int\limits_{\BR^{3n}}
\,d\vec x
\,d\vec y \,d\vec p {\Phi}^*(\vec x,t,\h)\times\\
&\times&\exp \Bigl(\frac i\h\lan\vec x -\vec y,\vec p \ran \Bigr)
[\De\vec p]^{\alpha_p}\biggl(\frac{\De \vec x+ \De\vec
y}{2}\biggr)^{\alpha_x} \Phi(\vec y,t,\h).
\end{eqnarray*}
Here
\[ \De \vec{y}=\vec y -\vec X(t,\h). \]
After a change of variables
\[ \De \vec{x} =\sqrt{\h} \vec\xi,\quad
\De \vec{y} =\sqrt{\h} \vec\zeta, \quad \De \vec{p} =\sqrt{\h}
\vec\omega \] and using the formula for  function $\Phi (\vec x,t,\h)$
 from the class $\CP^t_\h(Z(t,\h))$ (\ref{bbst1.5}), we find
\begin{eqnarray*}
\sigma_\alpha(t,\h)&=& \frac 1
{{(2\pi\h)}^n}{\h}^{{3n}/2}{\h}^{{|\alpha|}/2}2^{-|\alpha_x|}
\int\limits_{\BR^{3n}} d\vec \xi d\vec \zeta d\vec
\omega\va^*(\vec \xi, t,\h)
\times\\
&\times& \exp\{i \lan\vec \xi -\vec \zeta,\vec \omega \ran\}
\vec\omega\,{}^{\alpha_p}(\vec\xi+\vec\zeta)^{\alpha_x}\va(\vec\zeta,t,\h)=\\[8pt]
& =&\h^{(n+|\alpha|)/2}M_\alpha(t,\h),\\
{}\|\Phi\|^2&=&{\h}^{n/2} \int\limits_{\BR^n} d \vec \xi\va^*
(\vec \xi ,t,\h) \va (\vec \xi,t,\h)=\h^{n/2}M_0(t,\h).
\end{eqnarray*}
Recall that the function $\va(\vec\xi ,t,\h)$ depends on $\sqrt\h$
regularly, and $M_0(t,\h)>0$. Therefore \begin{eqnarray*} 
\De_\alpha(t,\h)=\dac{\sigma_\alpha(t,\h)}{\|\Phi\|^2}
=\h^{|\alpha|/2}\frac{M_\alpha(t,\h)}{M_0(t,\h)}
\leqslant\displaystyle\h^{|\alpha|/2}\max_{t \in {[0,T]}}
\frac{M_{\alpha}(t,\h)}{M_0(t,\h)}=O\big(\h^{|\alpha|/2}\big),
\end{eqnarray*}
Q.E.D.
\\

{\bf 2.} Proof of (\ref{bbst1.9}) follows from the explicit form
of a trajectory-coherent function $\Phi(\vec x,t,\h)\in\CP_\h^t$
(\ref{bbst1.5}) and the estimations (\ref{bbst1.8}).
\\

{\bf 3.}  Proof of (\ref{bbst1.11}).
Consider a function
$\phi(\vec{x})\in \mathbb{S}$. Then for any function $\Phi(\vec
x,t,\h)\in\CP_\h^t$ the integral
\[ \Big\lan\frac{|\Phi(t,\h)|^2}{\|\Phi(t,\h)\|^2}\Big|\,\phi\Big\ran=
\frac 1{\|\Phi(t,\h)\|^2}\inl_{\BR^n_x} \phi(\vec x)| \Phi(\vec
x,t,\h)|^2\,d\vec x=\frac 1{\|\va(t,\h)\|^2}\inl_{\BR^n_x}
\phi(\vec x)\Big|\va\Bigl(\frac{\De\vec
x}{\sqrt\h},t\Bigr)\Big|^2\,d\vec x \]  after the change of
variables $\vec\xi=\De\vec x/\sqrt\h$ becomes \[
\big\lan|\Phi(t,\h)|^2\big|\,\phi\big\ran=\frac{\h^{n/2}}{\|\va(t,\h)\|^2}
\inl_{\BR^n_\xi} \phi(\vec
X(t,\h)+\sqrt\h\vec\xi)|\va(\vec\xi,t,\h)|^2\, d\vec\xi. \]
Taking the limit $\h\to 0$ and  using that
\[ \|\va(t,\h)\|^2=\h^{n/2}\inl_{\BR^n_\xi} |\va(\vec\xi,t,\h)|^2\,
d\vec\xi, \]
where the function
$\va(\vec\xi,t,\h)$ depends on $\sqrt\h$ regularly, we obtain
the statement.
\\

 Proof of (\ref{bbst1.12}) is similar to the previous one, if we note that
the Fourier image of the function $\Phi(\vec x,t,\h)\in\CP_\h^t$
may be represented in the form \[ \tilde\Phi(\vec
p,t,\h)=\exp\Bigl\{\frac i\h\big[S(t,\h)- \lan \vec p,\vec
X(t,\h)\ran \big]\Bigr\}\tilde\va\Bigl( \frac{\vec p-\vec
P(t,\h)}{\sqrt\h},t,\h\Bigr), \] where \[
\tilde\va(\vec\omega,t,\h)=\frac 1{(2\pi)^{n/2}}\inl_{\BR^n_\xi}
e^{-i \lan\vec\omega, \vec\xi\ran }\va(\vec\xi,t,\h)d\xi. \]
\\


\end{document}